\documentclass[11pt,draft]{article}
\usepackage{amsfonts}
\usepackage{amssymb}
\usepackage{palatino}

\newlength{\extraspace}
\newlength{\extraspaces}
\newcommand{\B}[1]{{\mathbb #1}}
\newcommand{\C}[1]{{\mathcal #1}}

\newcommand{\beq}{\begin{equation}}
\newcommand{\eeq}{\end{equation}}
\newcommand{\beqn}{\begin{equation*}}
\newcommand{\eeqn}{\end{equation*}}
\newcommand{\bea}{\begin{eqnarray}}
\newcommand{\eea}{\end{eqnarray}}
\newcommand{\bean}{\begin{eqnarray*}}
\newcommand{\eean}{\end{eqnarray*}}
\newcommand{\nn}{\nonumber}

\newcommand{\half}{\frac 12}

\newcommand{\quarter}{\frac 14}
\newcommand{\eighth}{\frac 18}

\newcommand{\lish}{\langle\!\langle}
\newcommand{\rish}{\rangle\!\rangle}

\newcommand{\torus}[2]{{}^{#1}\mathop{\mbox{\LARGE$\square$}}_{#2}}
\newcommand{\Tr}{\mathop{\mbox{Tr}}}

\newcommand{\lb}{\lbrack}
\newcommand{\rb}{\rbrack}

\newcommand{\msc}[1]{\mbox{\scriptsize #1}}

\newcommand{\om}{\omega}
\newcommand{\al}{\alpha}

\newcommand{\ket}[1]{{\left|#1\right\rangle}}
\newcommand{\bra}[1]{{\left\langle#1\right|}}

\newcommand{\hket}[1]{\widehat{\left|#1\right\rangle}}
\newcommand{\hbra}[1]{\widehat{\left\langle#1\right|}}

\newcommand{\dket}[1]{{\left.\left|#1\right\rangle\right\rangle}}
\newcommand{\dbra}[1]{{\left\langle\left\langle#1\right|\right.}}
\newcommand{\hdket}[1]{\widehat{\left.\left|#1\right\rangle\right\rangle}}

\newcommand{\tL}{\bar{L}}
\newcommand{\tJ}{\bar{J}}

\newcommand{\Ad}{\mbox{Ad}}

\newcommand{\cA}{{\cal A}}

\newcommand{\cT}{{\cal T}}

\newcommand{\cN}{{\cal N}}

\newcommand{\cR}{{\cal R}}

\newcommand{\cH}{{\cal H}}

\newcommand{\ty}{\tilde{y}}
\newcommand{\tx}{\tilde{x}}

\newcommand{\hchi}{\widehat{\chi}}
\newcommand{\hZ}{\widehat{Z}}

\newcommand{\Th}[2]{\Theta_{#1,#2}}
\renewcommand{\th}{{\theta}}
\newcommand{\tTh}[2]{\widetilde{\Theta}_{#1,#2}}

\newcommand {\eqn}[1]{(\ref{#1})}

\makeatletter
\@addtoreset{equation}{section}
\def\theequation{\thesection.\arabic{equation}}
\makeatother

\begin{document}

\thispagestyle{empty}

%%% Date, Preprint numbers, etc.
\begin{flushright}
{UT}-07-24\\
{YITP}-07-50\\
{HIP}-07-42/TH
%{\tt arXiv:YYMM.NNNN[hep-th]}\\
%\today
%revised 24th June 2005\\
\end{flushright}
\vspace{.3cm}

\begin{center}
{\LARGE\bf{
% Title
%The title line 1\\[2mm] and line 2
D-branes in T-fold conformal field theory
} }\\[15mm]

{\large Shinsuke Kawai${}^{\dag\ddag}$}\footnote{{\tt shinsuke.kawai(AT)helsinki.fi}} and
{\large Yuji Sugawara${}^\S$}\footnote{{\tt sugawara(AT)hep-th.phys.s.u-tokyo.ac.jp}}
\\[5mm]
% and Co Author\footnote{e-mail: co.author@inst.ac.uk}} \\[5mm]
${}^\dag${\it Yukawa Institute for Theoretical Physics, 
Kyoto University, Kyoto 606-8502, Japan}\\ %[4mm]
%and\\[4mm]
${}^\ddag${\it Helsinki Institute of Physics, 
P.O.Box 64, University of Helsinki, Helsinki FIN-00014, Finland}\\
${}^\S${\it Department of Physics, University of Tokyo, 7-3-1 Hongo, 
Bunkyo-ku, Tokyo 113-0033, Japan}
\\[15mm]
{\bf Abstract}

\begin{center}
\begin{minipage}{14cm}
% Abstract

We investigate boundary dynamics of orbifold conformal field theory  
involving T-duality twists. Such models typically appear in contexts of
non-geometric string compactifications that are called  monodrofolds 
or T-folds  in recent literature.
We use the framework of boundary conformal field theory to analyse 
the models from a microscopic world-sheet perspective. 
In these backgrounds there are two kinds of D-branes that are 
analogous to bulk and fractional branes in standard orbifold models. 
The bulk D-branes in T-folds allow intuitive geometrical interpretations and are 
consistent with the classical analysis based on the doubled torus formalism. 
The fractional branes, on the other hand, are `non-geometric' at any point 
in the moduli space and 
%their geometric counterparts seem to be missing 
have not been considered 
in the doubled torus analysis so far.
We compute cylinder amplitudes between the bulk and fractional branes, and
find that the lightest modes of the open string spectra show intriguing non-linear
dependence on the moduli (location of the brane or value of the Wilson line), 
suggesting that the physics of T-folds, when D-branes are involved, could 
deviate from geometric backgrounds even at low energies.
We also extend our analysis to the models with $SU(2)$ WZW fibre at arbitrary levels.

\end{minipage}
\end{center}

\end{center}

\noindent

\vfill\noindent
% PACS and keywords
PACS number(s): 11.25.Hf, 11.25.Uv

\newpage
\tableofcontents

%\newpage
\pagestyle{plain}
\setcounter{page}{1}
\setcounter{footnote}{0}

%%%%%%%%%%%%%%%%%%%%%%%%%%%%%%%%%%%%%%%%%%%%%%%%%%%
%%%%%%%%%%%%%%%%%%%%%%%%%%%%%%%%%%%%%%%%%%%%%%%%%%%

%%%%%
\section{Introduction and summary}
%%%%%

Recently much attention has been focused on a class of string backgrounds that involve
duality twists\cite{Hull:2004in,Flournoy:2004vn,Dabholkar:2005ve,Dabholkar:2002sy,Hellerman:2002ax,Serone:2003sv}.
These backgrounds are formulated as fibrations over a base manifold in which the transition functions are built from discrete duality transformations over and above the standard continuous (diffeomorphism and gauge) transformations, so that the fibre picks up non-trivial monodromies as it goes around cycles on the base. 
As the dualities are no less fundamental symmetries of the theory than
the diffeo and gauge symmetries it is natural to suppose that these are
as good backgrounds for strings as standard manifold backgrounds
(`geometric backgrouds').
In recent literature such backgrounds are called `monodrofolds'\cite{Flournoy:2004vn} or, when the duality used in the construction is T-duality, `T-folds'\cite{Hull:2004in} in particular.
In the present paper we shall be concerned only with T-folds.

T-folds are an example of {\it non-geometric} backgrounds 
and have features that differ from ordinary manifold backgrounds.
For instance the metric and the Kalb-Ramond field are not defined globally since T-duality mixes these two.
For T-folds of $d$-torus fibrations over a base manifold $B$ there exists a very useful framework known as the doubled-torus formalism, developed in \cite{Hull:2004in}.
This is to construct from the original T-fold an enlarged space $T^d\otimes\widetilde{T}^d\otimes B$ where $\tilde T^d$ (with coordinates $\tilde X=X_L-X_R$) is T-dual to $T^d$ (with $X=X_L+X_R$).
In the enlarged space the T-duality group $O(d,d;{\B Z})$ acts linearly.
The doubled torus is geometric and is considered as the collection of all possible T-duals associated with a given T-fold.
A T-fold is obtained from the doubled torus by projecting out redundant degrees of freedom.
The choice of physical degrees of freedom is called {\em polarisation} in \cite{Hull:2004in}.
The equations of motion of a T-fold are recovered from the doubled torus using appropriate constraints; hence the doubled torus with appropriate polarisation and the original T-fold are equivalent at classical level.
Classical T-fold backgrounds are also related to Hitchin's generalised complex 
geometry\cite{Gates:1984nk,Hitchin:2004ut,Gualtieri:2003dx}.
See \cite{Hull:2006va,Grange:2006es} for recent studies.

In string theory the space-time arises, in principle, as a consequence of the string world-sheet dynamics.
In particular, when studying non-geometric backgrounds that are somewhat beyond our intuitive understanding of spacetime, the world-sheet theory is expected to provide rich information beyond the supergravity approximation.
The world-sheet of T-folds is known to be described by conformal field theory (CFT) of asymmetric orbifolds\cite{Narain:1986qm}.
These are subject of recent intensive study motivated by phenomenological interests, as they give rise to various models of non-supersymmetric string backgrounds with vanishing
\cite{Kachru:1998hd,Angelantonj:1999gm} or exponentially suppressed\cite{Harvey:1998rc} cosmological constant.
An elementary check of legitimacy of such CFT is whether the model preserves modular invariance at one-loop level.
In stark contrast to the symmetric cases the level-matching in asymmetric orbifolds is not automatic and 
the one-loop partition functions often fail to be modular invariant.
As observed in \cite{Kachru:1998hd} it is nevertheless possible to construct consistent models of asymmetric orbifold in which the modular invariance is recovered by cancellation of level mismatch.
The authors of \cite{Hellerman:2006tx} re-consider this issue in the context of T-folds.
We review these technical details in Sec. \ref{sec:TfoldCFT}.

D-branes are essential in studying various non-perturabative aspects of string backgrounds, such as dual gauge theory, meta-stable vacua, and string duality.
They can also be used as a probe to analyse the geometry of the background.
D-branes on T-fold backgrounds are constructed and analysed in the doubled-torus picture in 
\cite{Lawrence:2006ma}, where classical D-brane spectrum consistent with the $O(d,d;{\B Z})$ monodromy was found in the model of $T^d$ fibrations over $S^1$.
In the present paper we study D-branes in a simple model of T-fold in the framework of world-sheet orbifold CFT, which would be complementary to \cite{Lawrence:2006ma}.
There are earlier work on D-branes in (different models of) asymmetric orbifolds, 
see {\em e.g. } \cite{Brunner:1999fj,Gaberdiel:2002jr}.
Our findings are summarised as follows:\\

\noindent
{\bf 1.}
We analyse D-brane spectrum in the T-fold model of $S^1$ fibration over $S^1$ base.
There are D-branes (bulk branes) that have geometric counterparts in the doubled picture.
They are identified with those found in \cite{Lawrence:2006ma}.\\

\noindent
{\bf 2.}
Furthermore, we also find D-branes involving the twisted sector (fractional branes), which are 
expected but not concretely constructed in \cite{Lawrence:2006ma}.
Computing overlaps reveals that both bulk and fractional branes satisfy Cardy conditions and hence they coexist in the T-fold background.
We find the mass of an open string stretched 
between the bulk and fractional branes shows 
intriguing non-linear dependence on the moduli.\\

\noindent
{\bf 3.}
We extend the analysis to T-fold models with $SU(2)_k$ fibration over $S^1$ and find similar results.

The plan of the rest of the paper is as follows. 
In the next section we describe the $S^1$ over $S^1$ model of T-fold CFT by reviewing discussions of
\cite{Flournoy:2005xe,Hellerman:2006tx,Hackett-Jones:2006bp}.
In section 3 we discuss D-branes in this background;
we construct boundary states of bulk and fractional branes, check their modular consistency (Cardy conditions) and discuss their properties.
In section 4 we consider world-sheet fermions, 
and in section 5 we generalise our discussions to
models with $SU(2)$ Wess-Zumino-Witten (WZW) fibre, and conclude with some comments.
Summary of formulae as well as technical issues are relegated to 4 appendices.

Throughout  this paper we use the convention of $\alpha'=1$.

%%%%%
\section{The world-sheet CFT}
\label{sec:TfoldCFT}
%%%%%

The example of T-fold that we shall consider in this section and the next is a circle fibration over a base of another circle, with the transition function being the T-dualisation so that the fibre transforms into its T-dual as it moves around the base 
\cite{Flournoy:2005xe,Hellerman:2006tx,Hackett-Jones:2006bp} 
(also, Chap.18 of \cite{Polchinski:1998rr}).
We set the radius of the base circle to be $R$ 
and that of the fibre circle to be at self-dual: $R'=1$, so as to make it
possible to gauge the T-duality symmetry. 
The fibre and the base coordinates are respectively 
$X(z,\bar z)=X_L(z)+X_R(\bar z)$ and
$Y(z,\bar z)=Y_L(z)+Y_R(\bar z)$.
The T-dualised fibre coordinates are $\tilde X(z,\bar z)=X_L(z)-X_R(\bar z)$.
The T-fold is defined as an `interpolating orbifold' on the covering space 
$S^1_{1} \times S^1_{2R}$,
whose orbifold action is the T-duality transformation on the fibre
accompanied by   
the half shift along the base circle\footnote
{
It is shown in \cite{Hackett-Jones:2006bp} that the doubled formalism
\cite{Hull:2004in,Dabholkar:2005ve} 
(see also \cite{Hackett-Jones:2006bp,Berman:2007vi,Chowdhury:2007ba,Berman:2007xn})
may be used to obtain the same one-loop partition function of this T-fold model.
In this paper, however, we shall not use the doubled torus formalism.
}
:
\beq
Y\rightarrow Y+2\pi R.
\label{eqn:BaseShift}
\eeq
%%%%%%%%%%%%%%
%%%%%%%%%%%%%%
In \cite{Flournoy:2005xe} it is discussed that the naive T-duality action 
\beq
T: X=(X_L, X_R)\rightarrow \tilde X=(X_L, -X_R),
\label{eqn:naiveT}
\eeq
leads to difficulty in modular invariance of the one-loop partition function\footnote
{
The failure of modular invariance originates from 
treating  the naive $T$-operator \eqn{eqn:naiveT} as
an order 2 automorphism, which is not the case. 
Indeed, as we will see later, one can still construct a modular invariant partition
function of the interpolating orbifold based on $T\otimes \cT_{2\pi R}$, which 
has an order 16 orbifold structure
(the fact that we should have an order 16 orbifold originates from the
level mismatch $1/16$ in the twisted sector). 
See also \cite{Aoki:2004sm}. 
However, we shall concentrate on the `improved' T-duality operator $T'$ (or
$T''$) defined later, since it is truly an order 2 automorphism and
consistent with the locality of vertex operators. 
}.
A reasonable remedy for this is proposed in \cite{Hellerman:2006tx}, by implementing
an appropriate shift in $X_L$ that renders the partition function of the asymmetric orbifold into 
essentially that of a (modular invariant) symmetric orbifold.
Similar construction of various T-duality orbifolds is discussed already in 
%%%
\cite{Kachru:1998hd,Angelantonj:1999gm,Harvey:1998rc,Erler:1996zs,Blumenhagen:1998uf}.
%%%
In this section we review the computation 
of the modular invariant one-loop partition function.
The system has central charge $c=2$ and may be considered as a part of critical bosonic string theory.
We will not mention the other $c=24$ components below, however.

%%%
\subsection{Locality of vertex operators and T-duality}
%%%

Before discussing the partition function, 
we present the argument \cite{Hellerman:2006tx} on
how the T-duality should act on vertex operators in a way consistent
with locality. 
To make things simple we focus only on the fibre part.
Consider the vertex operator,
\beq
{\C V}_{k_L, k_R}(z,\bar z)=C_{k_L, k_R}:e^{ik_LX_L+ik_RX_R}:.
\label{eqn:FibreVertexOp}
\eeq
The cocycle factor $C_{k_L, k_R}$ is defined as\footnote
{
Our conventions follow \cite{Polchinski:1998rr}.
The authors of \cite{Hellerman:2006tx} use a different convention with an extra factor 
$\exp(-\half\pi inw)$ but the difference is not essential in subsequent discussions.
}
\beq
C_{k_L, k_R}\equiv e^{\pi i w\hat n}.
\label{eqn:cocycle1}
\eeq
We denote the momentum and winding number {\em operators} 
with hats $\hat n$, $\hat w$
to distinguish them from corresponding numbers (eigenvalues) $n$ and $w$. 
They are related to the left and right moving momentum operators by
\bea
&&\hat p_L=\hat n+ \hat w,~~~ \hat p_R=\hat n-\hat w.
\eea
The eigenvalues for the operators $\hat p_L$ and $\hat p_R$ are $k_L$ and $k_R$.
We use round brackets to write the vertex operator (\ref{eqn:FibreVertexOp}) in terms of a pair of integers $n$ and $w$ instead of $k_L$ and $k_R$, 
\beq
{\C V}_{(n,w)}(z,\bar z)={\C V}_{k_L, k_R}(z,\bar z).
\eeq
The cocycle has been included to make these vertex operators mutually local,
\beq
{\C V}_{k_L, k_R}(z,\bar z){\C V}_{k'_L, k'_R}(z',\bar z')
={\C V}_{k'_L, k'_R}(z',\bar z'){\C V}_{k_L, k_R}(z,\bar z).
\eeq
%%%%%%%%%%

The vertex operator dual to (\ref{eqn:FibreVertexOp}) 
under the (naive) T-operation \eqn{eqn:naiveT} would then be
\bea
\hspace{-1cm}
T~:~ {\C V}_{k_L, k_R}(z,\bar z)~ \longrightarrow ~ 
e^{\pi i w\hat w}:e^{i k_LX_L(z)-i k_R X_R(\bar z)}:
=e^{\pi i \tilde{n} \hat w}:e^{i\tilde{k}_L X_L(z)+i\tilde{k}_R X_R(\bar
z)}:
\equiv \widetilde{\C V}_{\tilde{k}_L,\tilde{k}_R}(z,\bar{z}).
\label{eqn:naiveVO}
\eea
Here $\tilde k_L\equiv n+w= k_L$, 
$\tilde k_R\equiv  w-n =-k_R$, and $\tilde{n}\equiv w$, $\tilde{w}\equiv
n$.
Note that the T-dualised cocycle factor appearing in (\ref{eqn:naiveVO}), 
$
\tilde C_{\tilde{k}_L, \tilde{k}_R}\equiv e^{\pi i \tilde n\hat w},
%\label{eqn:cocycle2}
$
differs from the original one (\ref{eqn:cocycle1}). 
The operators ${\C V}_{k_L, k_R}$ and $\widetilde{\C V}_{k'_L, k'_R}$ are 
{\it not} mutually local when $wn'+nw'\in 2{\B Z}+1$, 
as can be seen from their operator product 
\beq
{\C V}_{k_L, k_R}(z,\bar z)\widetilde{\C V}_{k'_L, k'_R}(z',\bar z')
=e^{\pi i(wn'+nw')}\widetilde{\C V}_{k'_L, k'_R}(z',\bar z')
{\C V}_{k_L, k_R}(z,\bar z).
\label{eqn:COP}
\eeq
This would not cause any problem were we dealing with two separate
theories that are T-dual to each other. 
In the case of T-fold, however, we encounter such cross operator products and their non-locality indicates inconsistency of the model; in order to construct a sensible model we need to make the product (\ref{eqn:COP}) local.
This can be accomplished by including the appropriate factor of $e^{\pi i\hat n\hat w}$
\cite{Hellerman:2006tx} into the definition of the T-duality transformation\footnote
{
In \cite{Berman:2007vi} relation between this factor and a topological term in the supergravity description is discussed.
}.
This `improved' T-transformation (which we shall denote by $T'$) 
acts on states as 
\beq
T': \vert n,w,N^i,\bar N^i\rangle \rightarrow (-1)^{\sum \bar N^i}
e^{i\pi \hat n\hat w}\vert w,n,N^i,\bar N^i\rangle,
\label{eqn:TonStates}
\eeq
where $N^i$ and $\bar N^i$ are the left and right occupation numbers.
For the vertex operators, this operates as
\beq
T': {\C V}_{(n,w)}(z,\bar z)%={\C V}^{R'}_{k_L, k_R}(z,\bar z) 
\rightarrow e^{-i\pi nw}e^{i\pi \tilde w \hat n} :e^{i\tilde k_L X_L+i
\tilde k_R X_R}:
=e^{-i\pi nw}{\C V}_{(w,n)}(z, \bar z).
\label{eqn:TV}
\eeq 
Thus the improved T-operator $T'$ acts on vertex operators as 
$n \, \leftrightarrow \, w$
{\em while keeping the cocycle factor $C$ unchanged} up to a C-number phase;
this assures the mutual locality of vertex operators.

%%%%
We also note that $T'$ is actually involutive, $(T')^2 =  {\bf 1}$, on the
whole Hilbert space, whereas $T$ is not. 
This is because $T$ is interpretable as operator $({\bf 1}, e^{i\pi \bar J^1_0})$ in terms of 
the $SU(2)_1$ current $J^a$ characterizing the self-dual compact boson
(note that $e^{2\pi i J^1_0} \neq {\bf 1}$; it generates a non-trivial phase).
%%%

%%%
\subsection{The T-fold as an orbifold}
%%%

We defined the world-sheet CFT of the T-fold as an asymmetric orbifold 
on the covering space 
%%%%%%%%%%%%%
%%%%%%%%%%%%%
$S^1_{1}\times S^1_{2R}$, with order 2 orbifolding group
%%%%%%%%%%%%%
%%%%%%%%%%%%%
$G=\{I, \sigma\}$ where $I$ is the identity 
and $\sigma$ is T-dualisation of the fibre combined with 
the half shift in (the covering space of) the base 
${\C T}_{2\pi R}$: $Y\rightarrow Y+2\pi R$.
The computation of the one-loop T-fold partition function 
then follows the standard theory of orbifold,
%\cite{Dijkgraaf:1989hb},
\beq
Z^{\mbox{\scriptsize T-fold}}(\tau,\bar\tau)
=\frac{1}{|G|}\sum_{g,h\in G} %\epsilon_{(g|h)} 
\torus{h}{g}(\tau,\bar\tau)
= \half\left(
\torus II +\torus\sigma I +\torus I\sigma +\torus\sigma\sigma \right).
\label{eqn:orbZ}
\eeq
As the Virasoro zero-modes are sums of the fibre and base parts
$L_0=L^{\rm fibre}_0+L^{\rm base}_0$,
$\overline{L}_0= \overline{L}^{\rm fibre}_0+\overline{L}^{\rm base}_0$,
the partition trace in each sector sector-wise splits into the base and fibre parts,
\beq
\torus hg (\tau,\bar\tau)
=\Tr_{{\C H}_g} h q^{L_0-\frac{1}{12}}\bar q^{\overline L_0-\frac{1}{12}}
=Z^{\rm base}_{[g,h]}(\tau,\bar\tau)Z^{\rm fibre}_{[g,h]}(\tau,\bar\tau), 
\label{eqn:PT}
\eeq
where
\bea
Z^{\rm base}_{[g,h]}&=&\Tr_{{\C H}^{\rm base}_g}h q^{L_0^{\rm base}-\frac{1}{24}}
\bar q^{\overline L_0^{\rm base}-\frac{1}{24}},
\label{eqn:baseZ}\\
Z^{\rm fibre}_{[g,h]}&=&\Tr_{{\C H}^{\rm fibre}_g} h q^{L_0^{\rm fibre}-\frac{1}{24}}
\bar q^{\overline L_0^{\rm fibre}-\frac{1}{24}}.
\label{eqn:fibreZ}
\eea
Here ${\C H}_I$ (${\C H}_{\sigma}$) 
is the Hilbert space of the untwisted (twisted) sector.

\subsection{The fibre part of the partition function}

Below we describe an explicit computation of (\ref{eqn:fibreZ}) in the operator (rather than path-integral) formalism.
This is essentially the modular orbit completion\cite{Dijkgraaf:1989hb,Aoki:2004sm} using the orbifolding group that has been spelled out in (\ref{eqn:TonStates}).
We first look at the untwisted Hilbert space with no twist insertion, 
$Z^{\rm fibre}_{[I,I]}$.
The Virasoro zero-modes in this  sector can be written using the number operators 
$\hat N_k=\frac 1k a_{-k}a_k$ and $\hat{\overline{N}}_k=\frac 1k\bar a_{-k}\bar a_k$
($a_k$ and $\bar a_k$ are the mode operators of $X_L$ and $X_R$) as
\bea
&&L^{{\rm fibre},U}_0=\sum_{k=1}^\infty k\hat N_k
+\frac{1}{4}\left(\hat n +\hat w\right)^2,\nn\\
&&\overline{L}^{{\rm fibre},U}_0=\sum_{k=1}^\infty k\hat{\overline{N}}_k
+\frac{1}{4}\left(\hat n -\hat w\right)^2.
\label{eqn:L0}
\eea
The Hilbert space ${\C H}^{\rm fibre}_I$ is
\beq
{\C H}^{\rm fibre}_I=
\bigoplus_{N_p,\bar N_q}\bigoplus_{n,w} 
a_{-1}^{N_1} a_{-2}^{N_2}\cdots\bar a_{-1}^{\bar N_1}\bar a_{-2}^{\bar N_2}\cdots\vert (n,w)\rangle,
%\nn\\
%&=&\bigotimes_{p,q=1}^{\infty}
%\left(\bigoplus_{N_p=0}^\infty a_{-p}^{N_p}\right)
%\left(\bigoplus_{\bar N_q=0}^\infty \bar a_{-q}^{\bar N_q}\right)
%\bigoplus_{n,m\in{\B Z}}\vert (n,w)\rangle.
\eeq
with $N_p$ and $\bar N_q$ non-negative integers and $n,w\in {\B Z}$.
Now using (\ref{eqn:L0}) and taking the trace over ${\C H}^{\rm fibre}_I$ one finds,
\bea
Z^{\rm fibre}_{[I,I]}(\tau,\bar\tau)
&=&\frac{1}{|\eta(\tau)|^2}
\sum_{n,w\in{\B Z}}
\langle (n,w)\vert q^{\frac{1}{4}(\hat n+\hat w)^2}
\bar q^{\frac{1}{4}(\hat n-\hat w)^2}\vert (n,w)\rangle\nn\\
&=&\frac{1}{|\eta(\tau)|^2}\sum_{n,w\in{\B Z}}q^{\quarter(n+w)^2}\bar q^{\quarter(n-w)^2}
=\left |\frac{\theta_2(2\tau)}{\eta(\tau)}\right |^2+\left |\frac{\theta_3(2\tau)}{\eta(\tau)}\right |^2.
\label{eqn:IItorus}
\eea
For computing $Z^{\rm fibre}_{[I,\sigma]}=Z^{\rm fibre}_{[I,T']}$ we split the untwisted space 
${\C H}^{\rm fibre}_I$ into T-even and T-odd subspaces, 
\bea
{\C F}_+&=&\bigoplus_{\stackrel{N_p,\bar N_q}{\scriptstyle \sum \bar N_q={\rm even}}}
\bigoplus_{n,w} a_{-1}^{N_1} a_{-2}^{N_2}\cdots\bar a_{-1}^{\bar N_1}\bar a_{-2}^{\bar N_2}\cdots
(\vert (n,w)\rangle+(-1)^{nw}\vert (w,n)\rangle)\nn\\
&&\oplus\bigoplus_{\stackrel{N_p,\bar N_q}{\scriptstyle \sum \bar N_q={\rm odd}}}
\bigoplus_{n,w} a_{-1}^{N_1} a_{-2}^{N_2}\cdots\bar a_{-1}^{\bar N_1}\bar a_{-2}^{\bar N_2}\cdots
(\vert (n,w)\rangle-(-1)^{nw}\vert (w,n)\rangle),\nn\\
{\C F}_-&=&\bigoplus_{\stackrel{N_p,\bar N_q}{\scriptstyle \sum \bar N_q={\rm odd}}}
\bigoplus_{n,w} a_{-1}^{N_1} a_{-2}^{N_2}\cdots\bar a_{-1}^{\bar N_1}\bar a_{-2}^{\bar N_2}\cdots
(\vert (n,w)\rangle+(-1)^{nw}\vert (w,n)\rangle)\nn\\
&&\oplus\bigoplus_{\stackrel{N_p,\bar N_q}{\scriptstyle \sum \bar N_q={\rm even}}}
\bigoplus_{n,w} a_{-1}^{N_1} a_{-2}^{N_2}\cdots\bar a_{-1}^{\bar N_1}\bar a_{-2}^{\bar N_2}\cdots
(\vert (n,w)\rangle-(-1)^{nw}\vert (w,n)\rangle).
\eea
It can be checked that $T'u=\pm u$, $u\in{\C F}_\pm$.
Taking the trace over ${\C H}^{\rm fibre}_I$ with $T'$ inserted in the temporal direction,
we see that when $n\neq w$ the traces over ${\C F}_+$ and ${\C F}_-$ cancel each other, so the contribution comes only from the fixed points $n=w$ of the $T'$-transformation. 
The trace is then,
\bea
Z^{\rm fibre}_{[I,\sigma]}(\tau,\bar\tau)
&\equiv&\Tr_{{\C H}^{\rm fiber}_I} 
T' q^{L_0^{\rm fiber}-\frac{1}{24}}\bar q^{\overline L_0^{\rm fiber}-\frac{1}{24}}\nn\\
&=&\sum_{\stackrel{\scriptstyle n=w}{n,w\in{\B Z}}}
\langle (n,w)\vert \frac{(-1)^{nw}q^{\frac{1}{4}(\hat n+\hat w)^2}
\bar q^{\frac{1}{4}(\hat n-\hat w)^2}}{\eta(\tau)
\bar q^{\frac{1}{24}}\prod_{k=1}^\infty(1+\bar q^k)}\vert (w,n)\rangle\nn\\
%&=&\frac{1}{\eta(\tau)}\sqrt{\frac{2\bar\eta(\bar\tau)}{\bar\theta_2(\bar\tau)}}
%\sum_{n\in{\B Z}}
%(-1)^nq^{\frac{\alpha'}{4}(\frac{1}{R'}+\frac{R'}{\alpha'})^2n^2}
%\bar q^{\frac{\alpha'}{4}(\frac{1}{R'}-\frac{R'}{\alpha'})^2n^2}.
&=&  
\frac{1}{\eta(\tau)}\sum_{n \in {\B Z}} (-1)^n q^{n^2} \cdot 
\overline{\sqrt{\frac{2\eta(\tau)}{\theta_2(\tau)}}}
%\nn\\
%&=&
= \left| \frac{2\eta(\tau)}{\theta_2(\tau)}\right|.
\label{eqn:TItorus}
\eea
In the last line we made use of identity \eqn{useful identity}.
Taking modular transformations we also obtain
\bea
Z^{\rm fibre}_{[\sigma,I]}(\tau,\bar\tau)
&\equiv &\Tr_{{\C H}^{\rm fibre}_\sigma} q^{L_0^{\rm fibre}-\frac{1}{24}}
\bar q^{\overline L_0^{\rm fibre}-\frac{1}{24}}
\, \left(=Z^{\rm fibre}_{[I,\sigma]}(-1/\tau,-1/\bar\tau) \right)\,
=\left| \frac{2\eta(\tau)}{\theta_4(\tau)}\right|,
\label{eqn:ITtorus}\\
Z^{\rm fibre}_{[\sigma,\sigma]}(\tau,\bar\tau)
&\equiv &\Tr_{{\C H}^{\rm fibre}_\sigma} T'q^{L_0^{\rm fibre}-\frac{1}{24}}
\bar q^{\overline L_0^{\rm fibre}-\frac{1}{24}}
\, \left(
=Z^{\rm fibre}_{[\sigma,I]}(\tau+1,\bar\tau+1) \right)\,
=\left| \frac{2\eta(\tau)}{\theta_3(\tau)}\right|.
\label{eqn:TTtorus}
\eea
The expressions (\ref{eqn:IItorus}), (\ref{eqn:TItorus}), (\ref{eqn:ITtorus}), (\ref{eqn:TTtorus}) are the
the partition traces of the fibre part of the T-fold.
They are nothing but those of the $c=1$ CFT 
at the Kosterlitz-Thouless point.

%%% An alternative description: the shift/reflection chiral orbifolds %%%

In the computations above it was essential to include in the definition of T-duality (\ref{eqn:TonStates}) the phase factor $e^{i\pi\hat n\hat w}$ that is associated with the locality of vertex operators. 
Since $\hat n\hat w=\quarter (\hat p_L^2-\hat p_R^2)$ this factor contributes 
$e^{\frac{i\pi}{4}\hat p_L^2}$ to the left-moving sector and
$e^{-\frac{i\pi}{4}\hat p_R^2}$ to the right-moving sector.
The authors of \cite{Hellerman:2006tx} also compute the same partition traces based on a slightly different approach, with T-duality defined by
\beq
T'': X_L\rightarrow X_L+\half\pi,\;\;\;
X_R\rightarrow -X_R,
%Y\rightarrow Y+2\pi R.
\label{eqn:T2}
\eeq
instead of (\ref{eqn:TonStates})
\footnote
{
The vertex operators are transformed under $T''$ as
$
e^{ik_LX_L}\rightarrow e^{\frac{i\pi}{2}(n+w)}e^{ik_LX_L}$, $
e^{ik_RX_R}\rightarrow e^{-ik_RX_R},
$
so at the fixed points $n=w$ of the orbifold the shift in $X_L$ yields the same phase factor as $e^{i\pi nw}$ of (\ref{eqn:TonStates}), giving the same contribution to the partition trace as in (\ref{eqn:TItorus}).
When $n\neq w$, $T'$ and $T''$ generate different phases in vertex operators.
It is argued in \cite{Hellerman:2006tx} that the difference of the phase
factor can be absorbed into the normalization of the ground states.
}.
A merit of this approach is that the left and right sectors of the fibre 
can be treated separately as two chiral orbifolds 
whose covering spaces are both $S^1$ at self-dual radius.
The left part of the action (\ref{eqn:T2}) 
generates a shift orbifold, 
namely CFT of a boson on $S^1$ at the radius reduced by half
({\em i.e.} it operates as a `chiral half-shift operator').
The right part generates a reflection orbifold $S^1/{\B Z}_2$, i.e. a line element of length $\pi$. 
As is well known these two chiral CFTs are equivalent;
the orbifold group (\ref{eqn:T2}) acts on currents 
$J^{\pm}=e^{\pm 2i X_{L}}$, $J^3=i\partial X_{L}$,
$\tJ^{\pm}=e^{\pm 2i X_{R}}$, $\tJ^3=i\partial X_{R}$,
of the underlying $SU(2)_L\times SU(2)_R$ 
symmetry as
$J^\pm \rightarrow -J^\pm $, $J^3 \rightarrow J^3$, 
$\tJ^\pm\rightarrow \tJ^\mp$, $\tJ^3\rightarrow-\tJ^3$, or
\bea
&&J^1 \rightarrow -J^1 ,\;\;\; J^2\rightarrow -J^2,\;\;\; J^3\rightarrow J^3,\nn\\
&&\tJ^1\rightarrow \tJ^1,\;\;\; \tJ^2\rightarrow -\tJ^2,\;\;\; \tJ^3
\rightarrow -\tJ^3.
\label{eqn:TonJs}
\eea
In other words one can identify
\begin{eqnarray}
 T'' = (e^{i\pi J^3_0}, e^{i \pi \bar J^1_0}).
\label{T''}
\end{eqnarray}
As the left and right actions of $T''$ are equivalent up to a global $SU(2)$ rotation, 
the resulting orbifold CFTs should be equivalent.

In this picture the left and right CFTs are represented 
by chiral bosons with (anti-)periodic boundary conditions,
\beq
X_L(z+k\omega_1+\ell\omega_2)=X_L(z)+\half\pi(k(2w+\alpha)+\ell(2m+\beta)),
\label{eqn:FL}
\eeq
\beq
X_R(\bar z+k\bar\omega_1+\ell\bar\omega_2)=
\left\{
\begin{array}{cc}
X_R(\bar z)+\pi(k w+\ell m) & (\alpha,\beta)=(0,0), \\
e^{\pi i(k\alpha+\ell\beta)}X_R(\bar z) & (\alpha,\beta)\neq(0,0),
\end{array}\right.
\label{eqn:FR}
\eeq
where $\alpha, \beta\in\{ 0,1\}$ represent boundary conditions and correspond to  
$0\leftrightarrow I$ and $1\leftrightarrow\sigma$ of the orbifold sectors.
$\omega_1(=1)$, $\omega_2(=\tau)$ are 
the two periods of the world-sheet torus and $w, m \in{\B Z}$.
The partition traces of these chiral bosons can be found by path-integral 
(see e.g. \cite{Itzykson:1986pj,DiFrancesco:1997nk}) 
and are shown to 
coincide with (\ref{eqn:IItorus}), 
(\ref{eqn:TItorus}), (\ref{eqn:ITtorus}), (\ref{eqn:TTtorus}).

In the following sections, we shall work with the $T''$-operator rather than
$T'$ in order to make the $SU(2)$-structure manifest.

%%%%%%%%%%%%%%%%%%%%%%%%%%%%%%%%%%%%%%%%%%%%%%%%%%%

%%%
\subsection{Modular invariance of the partition function}
%%%

The base part of the T-fold is a free boson $Y(z,\bar z)=Y_L(z)+Y_R(\bar z)$, defined (in the covering space) on $S^1$ of radius $2R$.
As the group action $\sigma$ of the orbifold shifts $Y$ by $2\pi R$, the periodicity of $Y$ is odd (even) integer multiple of $2\pi R$ when there is (there is not) a $\sigma$-twisting.
We thus consider periodic boundary conditions
\beq
Y_L(z+k\omega_1+\ell\omega_2)=Y_L(z)+\pi R(k w+\ell m),
\eeq
and likewise for $Y_R$, where $\omega_{1,2}$ are as in (\ref{eqn:FL}) and $k,\ell\in{\B Z}$. 
The partition function for each boundary condition $(w,m)$ is
\beq
Z_{R, (w,m)}(\tau,\bar\tau)
=\frac{R}{\sqrt{{\rm Im} \tau}}\frac{1}{\vert\eta(\tau)\vert^2}\exp\left\{
-\frac{\pi R^2\vert w\tau+ m \vert^2}{{\rm Im}\tau}\right\}.
\label{eqn:bosonZ}
\eeq
On the world-sheet $\omega_1$ ($\omega_2$) is the spatial (temporal)
direction as before. 
%with respect to the Hamiltonian $L_0+\overline L_0$.
As the twisting by ${\C T}_{2\pi R}$ in the $\omega_1$ ($\omega_2$) direction corresponds to $w$ ($m$) being odd, the partition trace (\ref{eqn:baseZ}) of each sector is obtained by summing up $w$ and $m$ of appropriate parities,
\bea
Z^{\rm base}_{[\alpha,\beta]}(\tau,\bar\tau)
&=& 
\sum_{w',m'\in {\B Z}} 
Z_{2R,(w'+\frac{\al}{2},m'+\frac{\beta}{2})}(\tau,\bar{\tau})
\left(\equiv 
2 \sum_{\stackrel{\scriptstyle w\in 2{\B Z}+\alpha}{m\in 2{\B Z}+\beta}}
Z_{R,(w,m)}(\tau,\bar\tau) \right)\nn\\
&=&
\frac{1}{\vert\eta(\tau)\vert^2}\sum_{k,\ell\in {\B Z}}(-1)^{\beta k}
q^{(\frac{k}{4R}+\frac{(2\ell+\alpha)R}{2})^2}
\bar q^{(\frac{k}{4R}-\frac{(2\ell+\alpha)R}{2})^2},
\label{eqn:baseZ2}
\eea
where $\alpha,\beta\in {\B Z}_2$, and we have Poisson-resummed to go to the last line.
The correspondence between the notation here and that of (\ref{eqn:baseZ}) is 
$(0,1)\leftrightarrow(I,\sigma)$.
As can be easily checked these partition traces are modular covariant:
%%%%%%%%%%%%%%%%%%%%%%%%%%%%%%%%%%%%%%%%%%%%%%%%%%%%%%%%%%%%%%%%%%%%%
\begin{eqnarray}
 Z^{\rm base}_{[\al,\beta]}(\tau+1,\bar{\tau}+1)&=& 
 Z^{\rm base}_{[\al,\al+\beta]}(\tau,\bar{\tau}), \nn\\
 Z^{\rm base}_{[\al,\beta]}(-1/\tau,-1/\bar{\tau})&=& 
 Z^{\rm base}_{[\beta,\al]}(\tau,\bar{\tau}).
\label{eqn:BaseModCov}
\end{eqnarray}
%%%%%%%%%%%%%%%%%%%%%%%%%%%%%%%%%%%%%%%%%%%%%%%%%%%%%%%%%%%%%%%%%%%%%
Assembling the base and the fibre pieces from the last subsection the
one-loop partition function of the T-fold reads
\beq
Z^{\mbox{\scriptsize T-fold}}(\tau,\bar\tau)
= \half Z^{\rm base}_{[0,0]}
\left(\left |\frac{\theta_2(2\tau)}{\eta(\tau)}\right |^2+\left |\frac{\theta_3(2\tau)}{\eta(\tau)}\right |^2\right)
+Z^{\rm base}_{[0,1]}\left| \frac{\eta(\tau)}{\theta_2(\tau)}\right|
+Z^{\rm base}_{[1,0]}\left| \frac{\eta(\tau)}{\theta_4(\tau)}\right|
+Z^{\rm base}_{[1,1]}\left| \frac{\eta(\tau)}{\theta_3(\tau)}\right|,
\label{eqn:fullZ}
\eeq
with $Z^{\rm base}_{[\alpha,\beta]}$ given by (\ref{eqn:baseZ2}).
As the fibre and the base parts are both modular covariant, the T-fold
partition traces (\ref{eqn:PT}) are modular covariant and hence the
partition function is modular invariant. 
%%%%
%%%%
Actually, \eqn{eqn:fullZ} is just the same partition function as that of 
the symmetric orbifold 
$$
\left\lb S^1_{1} \times S^1_{2R}\right\rb/ \Big(\cR \otimes \cT_{2\pi R}\Big),
$$
where $\cR$ acts as reflection on the fiber coordinates, 
$\cR\,:\,(X_L,X_R)\,\rightarrow\, (-X_L,-X_R)$.
This of course is expected from the above construction of the modular invariant. 
This does not mean, however, that the T-fold CFT (the asymmetric orbifold) 
describes the same physics as the symmetric orbifold. 
As we shall see in the next section, the physics of D-branes in these two models differs significantly;
this is one of our motivations to elaborate on the dynamics of T-fold boundary states in the next section.

We comment on T-duality along the base circle. 
The standard T-duality along the base is not a symmetry of the T-fold since the $U(1)$ 
isometry is broken by the orbifold construction. 
Instead, the following interpolating orbifold may be regarded as the T-dual of the T-fold along the base:
\begin{eqnarray}
 \left\lb S^1_1 \times S^1_{R/2}\right\rb/ 
 \Big(T'' \otimes \widetilde{\cT}_{2\pi \frac{1}{R}}\Big),
\label{T-dualized T-fold}
\end{eqnarray}
where the `dual translation' $\widetilde{\cT}_{2\pi \frac{1}{R}}$ is defined to act as 
$(Y_L,Y_R)\,\rightarrow\, (Y_L+2\pi \frac{1}{2R}, Y_R-2\pi\frac{1}{2R})$.
Note that $\widetilde{\cT}_{2\pi \frac{1}{R}}$ is interpretable as
the double covering operator 
$S^1_{R/2}/ \widetilde{\cT}_{2\pi \frac{1}{R}} \cong S^1_R$, 
corresponding precisely to the T-dual of the half-shift operator. 
The modular invariant of this model is computed to be
\beq
 Z^{\msc{T-dual. T-fold}}(\tau,\bar{\tau}) = \frac{1}{2}
\sum_{\al,\beta\in {\B Z}_2}
\widetilde{Z}^{\rm base}_{R, [\al,\beta]}(\tau,\bar{\tau})
Z^{\rm fibre}_{[\al,\beta]}(\tau,\bar{\tau}),
\label{Z T-dualized T-fold}
\eeq
%%%
where
\beq
\widetilde{Z}^{\rm base}_{R, [\al,\beta]}(\tau,\bar{\tau})\equiv
\sum_{w,m\in {\B Z}} (-1)^{\al m+ \beta w} Z_{R/2, (w,m)}(\tau,\bar{\tau}).
\eeq
For the dual radius $\tilde{R}= 1/R$ one can use the Poisson resummation to check that
\begin{eqnarray}
 && Z^{\rm base}_{R,[\al,\beta]}(\tau,\bar{\tau})
=\widetilde{Z}^{\rm base}_{\tilde{R},[\al,\beta]}(\tau,\bar{\tau}).
\end{eqnarray}
Hence the model (\ref{T-dualized T-fold}) indeed 
has the partition function equal to that of the original T-fold 
with the base $S^1$ at the dual radius.

%%%%%%%%%%%%%%%%%%%%%%%%%%%%%%%%%%%%%%%%%%%%%%%%%%%
%%%%%%%%%%%%%%%%%%%%%%%%%%%%%%%%%%%%%%%%%%%%%%%%%%%

%%%%%
\section{D-branes in the T-fold}
%%%%%

In this section
we study boundary states describing 
D-branes in the T-fold background described above. 
In orbifold theory there are two types of D-branes in general: 
{\em bulk} and {\em fractional} branes. 
The bulk branes are given by making the orbifold projection on the 
boundary states in the parent theory that are not invariant 
under the action of the orbifold group. In other words, these are just 
superpositions of branes and their `images' of the orbifold action. 
In the T-fold these roughly correspond to  
superposition of Dirichlet and Neumann states in the fibre, times a base state. 
%%%%
On the other hand, the fractional branes correspond to 
boundary conditions invariant under the orbifold action 
already in the parent theory 
(typically, the branes localized at the fixed points of orbifolds).
Their boundary states involve contributions from the twisted sectors
that are necessary for an orbifold projection in the open string 
Hilbert space \cite{Diaconescu:1999dt}. 
It turns out that the both types of branes exist in the T-fold model.
%%%

%%%
\subsection{Boundary conditions and boundary states}
%%%

We start with general remarks before constructing the boundary states.
The machinery of boundary conformal field theory 
is well developed for (symmetric) orbifold models 
\cite{Oshikawa:1996dj,Fuchs:1998xx,Birke:1999ik}.
Conformal field theory may generally have 
larger symmetries than Virasoro and the question 
of finding boundary states is closely related to which sub-symmetry of the full bulk symmetry the boundary should preserve.
Clearly the most elementary boundary states are the Virasoro boundary
states that are spanned by Virasoro Ishibashi states \cite{Ishibashi:1988kg}, 
since the conformal symmetry must be preserved by any boundary of CFT. 
In $c=1$ conformal theory there are other symmetries such as $U(1)$ or the enhanced symmetries 
${\C A}_N$ or ${\C A}_N/{\B Z}_2$ that are present at various special
points in the moduli space
\footnote{
The symmetries ${\C A}_N$ and ${\C A}_N/{\B Z}_2$ are cousins of the $SU(2)$ away from the self-dual point (in fact ${\C A}_1=SU(2)$). 
Notations of these rational models are summarised in Appendix \ref{sec:RCFT}.
${\C A}_4\simeq{\C A}_1/{\B Z}_2$ concerns us in studying the T-fold.}.
The boundary does not necessarily preserve such an extended chiral symmetry but when it does it carries corresponding charges of the symmetry. 
In the case of Virasoro the boundary carries the label of Virasoro weight.
For $U(1)$ the boundary is characterised by momenta and winding numbers. 
When the conserved symmetry is the extended symmetry ${\C A}_N$ or ${\C A}_N/{\B Z}_2$ the boundary is characterised by the representation labels of the rational CFT.
When the model allows a free field representation we also have familiar Dirichlet or Neumann states; Dirichlet is characterised by the position of the D-brane while a Neumann boundary can carry a Wilson line parameter.
In general we have better control of boundary states when the preserved symmetry is larger.
As we are ultimately interested in the physics of the string background we shall try to construct analogues of Dirichlet and Neumann states.
This is straightforward in the bulk brane case as the concept of Dirichlet and Neumann is just inherited from the parent theory. 
In constructing fractional states we will first look at the extended symmetries 
${\C A}_4$ and ${\C A}_1/{\B Z}_2$.

%%%%%%%%%%%%%%%%%%%%%%%%%%%%%%%%%%%%%%%%%%%%%%%%%%%%%%%%%%%%%%%%%%%

For constructed boundary states we shall check the Cardy conditions,
considering the cylindrical (annular) world-sheet. Namely, the closed string 
amplitude $Z^{c}(is)=\langle B_a|e^{-\pi s H^{c}}| B_b\rangle$ 
should be equated to  the open string one-loop amplitude 
$Z^{o}(it)=\Tr_{\cH_{ab}} e^{-2 \pi t H^{o}}$ by modular transformation 
$t=1/s$, with boundary conditions corresponding to
the boundary states $\bra{B_a}$ and $\ket{B_b}$. 
Here $H^c \equiv L_0+\tL_0-\frac{c}{12}$, 
$H^o\equiv L^{\msc{open}}_0- \frac{c}{24}$ 
are the closed and open string Hamiltonians. 
When both $\bra{B_a}$ and $\ket{B_b}$ are fractional branes, 
the open string amplitude has to be suitably 
orbifold-projected:
$Z^{o}(it)=\frac{1}{|G|} \sum_{h\in G}\Tr_{\cH_{ab}} \left\lb 
h e^{-2 \pi t H^{o}} \right\rb$.

%%%%%%%%%%%%%%%%%%%%%%%%%%%%%%%%%%%%%%%%%%%%%%%%%%%%%%%%%%%%%%%%%%

%%%
\subsection{Bulk branes}
\label{sec:bulkB}
%%%

Let us first recall that familiar Dirichlet and Neumann states of a compact boson on a circle of radius $R$ are
\bea
&&\vert D(x_0)\rangle_R= \frac{1}{2^{1/4}\sqrt{R}}
\sum_{n\in{\B Z}}e^{-i n x_0/R}\exp\left\{\sum_{k=1}^\infty\frac{1}{k}a_{-k}\bar a_{-k}\right\}
\vert(n,0)\rangle,\nn\\
&&\vert N(\tilde x_0)\rangle_R=\frac{\sqrt{R}}{2^{1/4}}
\sum_{w\in{\B Z}}e^{-iw\tilde x_0 R}\exp\left\{-\sum_{k=1}^\infty\frac{1}{k}a_{-k}\bar a_{-k}\right\}
\vert(0,w)\rangle,
\eea
with $x_0$ ($\tilde x_0$) parametrising the position of the D-brane 
(Wilson line on the Neumann state).
Their overall normalisation has been chosen so that the overlaps 
yield consistent open string spectra (the Cardy conditions)
($\Delta x_0\equiv x_0-x'_0$, $\Delta\tilde x_0\equiv \tilde x_0-\tilde
x'_0$,
$t\equiv 1/s$);
\bea
{}_R\langle D(x_0)\vert e^{-\pi sH^{c}}\vert D(x'_0)\rangle_{R}
&=&\frac{1}{\sqrt{2}R}\frac{1}{\eta(is)}
\sum_{n\in {\B Z}} e^{-2\pi s \frac{n^2}{4R^2}} e^{i\frac{\Delta x_0}{R}n} \nn\\
%\theta_3(\frac{\Delta x_0}{2\pi R}\vert\frac{is}{2R^2})\nn\\
&=&%\frac{1}{\eta(\tau)}q^{\frac{(\Delta x_0)^2}{4\pi^2\alpha'}}
%\theta_3(\frac{R\Delta x_0\tau}{\pi\alpha'}\vert\frac{2R^2\tau}{\alpha'})
\frac{1}{\eta(it)}\sum_{w\in{\B Z}}e^{-2\pi t(Rw+\frac{\Delta x_0}{2\pi})^2}
\equiv  Z_R^{DD}(it;\Delta x_0)
\nn\\
{}_R\langle N(\tilde x_0)\vert e^{-\pi sH^{c}}\vert N(\tilde x'_0)\rangle_R
&=&\frac{R}{\sqrt{2}}\frac{1}{\eta(is)}
\sum_{w\in {\B Z}} e^{-2\pi s \frac{R^2 w^2}{4}} e^{iR \Delta \tx_0 w} \nn\\
%\theta_3(\frac{R\Delta\tilde x_0}{2\pi}\vert\frac{is R^2}{2})\nn\\
&=&%\frac{1}{\eta(\tau)}q^{\frac{(\Delta\tilde x_0)^2}{4\pi^2\alpha'}}
%\theta_3(\frac{\Delta\tilde x_0\tau}{\pi R}\vert\frac{2\alpha'\tau}{R^2})
\frac{1}{\eta(it)}\sum_{n\in{\B Z}}e^{-2\pi t
(\frac{n}{R}+\frac{\Delta \tilde x_0}{2\pi})^2}
\equiv  Z_R^{NN}(it;\Delta\tilde x_0)
\nn\\
{}_R\langle D(x_0)\vert e^{-\pi sH^{c}}\vert N(\tilde x_0)\rangle_R
&=& \frac{1}{\sqrt{2}} \sqrt{\frac{2 \eta (is)}{\th_2(is)}}
%&=&\frac{1}{\sqrt 2}\frac{\theta_4(0\vert 2 is)}{\eta(is)}\nn\\
%&=&%\frac{\theta_2(0\vert\half\tau)}{2\eta(\tau)}=
%\frac{1}{\eta(it)}\sum_{n=0}^\infty e^{-2\pi t \quarter(n+\half)^2}
= \sqrt{\frac{\eta(it)}{\th_4(it)}}
\equiv Z^{DN}(it).
\label{eqn:overlaps}
\eea

We wish to find boundary states of 
the T-fold that are combination of such Dirichlet and Neumann states.
%%%
As already addressed, we regard the T-fold as the orbifold of 
$S^1_{1} \, \mbox{(fibre)} \times S^1_{2R} \, \mbox{(base)}$ with
respect to the involution $\sigma \equiv T'' \otimes \cT_{2\pi R}$,
where the improved $T$-operator $T''$ is defined by \eqn{eqn:T2}.

The position of a localized D-brane in the base
direction will be denoted by $y_0$,
and for a Neumann state the value of Wilson line in 
the base by $\tilde y_0$. On the other hand, in the fibre direction, 
it is convenient to express the open string modulus 
(position or Wilson line) by a common angle variable 
$\theta$ because the fibre circle is self-dual.

An obvious way of constructing a bulk brane is to act the T-fold
operator $\sigma \equiv T'' \otimes \cT_{2\pi R}$ on a boundary state
of $S^1_1 \times S^1_{2R}$. 
For instance, if taking Dirichlet conditions in both fibre and base directions,
the desired boundary state will be
\beq
\vert D(\theta)D(y_0)\rangle
=\frac{1+\sigma}{\sqrt 2}\vert D(\theta)\rangle^{\rm fibre}_{1}\otimes
\vert D(y_0)\rangle^{\rm base}_{2R}.
\label{eqn:bulkDD}
\eeq
The normalisation factor of $1/\sqrt 2$ is 
for consistency with the Cardy conditions (note that
$\half (1+\sigma)^2=1+\sigma$).
$\sigma$ acts on the base Dirichlet state as translation by $2\pi R$, 
\beq
\sigma:\vert D(y_0)\rangle^{\rm base}_{2R}
\rightarrow\vert D(y_0+2\pi R)\rangle^{\rm base}_{2R}
\eeq
while it acts trivially on the Neumann state,
\beq
\sigma:\vert N(\tilde y_0)\rangle^{\rm base}_{2R}
\rightarrow\vert N(\tilde y_0)\rangle^{\rm base}_{2R}.
\eeq
%%%
The action of $\sigma$ on the fibre is slightly non-trivial due to
phase ambiguity of the Fock vacua.
We choose the phase 
%and the moduli parameters 
so that $\sigma$ acts on
the fibre states as\footnote
{
The relation (\ref{T'' Fock vacua}) is not possible with the naive T-operation $T$, 
since $T^2\neq {\bf 1}$ and the extra phase cannot be absorbed
into normalisation of the Fock vacua.}
\beq
\sigma:\vert D(\th)\rangle^{\rm fibre}\leftrightarrow
\vert N(\th)\rangle^{\rm fibre},
\label{T'' Fock vacua}
\eeq
in accordance with the standard order 2 relation of T-duality $(T'')^2={\bf 1}$.

The bulk $DD$ brane (\ref{eqn:bulkDD}) 
is organised into a superposition of 
direct products of ordinary Dirichlet/Neumann states,
%\begin{subequations}% ***Requires amsmath package***
\beq
\vert D(\th)D(y_0)\rangle
=\frac{1}{\sqrt 2}\left(
\vert D(\th)\rangle_1^{\rm fibre}\otimes\vert D(y_0)\rangle^{\rm base}_{2R}
+\vert N(\th)\rangle_1^{\rm fibre}\otimes\vert D(y_0+2\pi R)\rangle^{\rm base}_{2R}
\right).
\label{eqn:bulkDD-2}
\eeq
One may construct similar states by starting from 
the $DN$, $ND$, $NN$ states and then 
projecting onto the invariant subspaces,
\bea
\vert D(\th)N(\tilde y_0)\rangle
&=&\frac{1+\sigma}{\sqrt 2}\vert D(\th)\rangle_1^{\rm fibre}\otimes
\vert N(\tilde y_0)\rangle^{\rm base}_{2R}\nn\\
&=&\frac{1}{\sqrt 2}\left(
\vert D(\th)\rangle_1^{\rm fibre}%\otimes\vert N(\tilde y_0)\rangle^{\rm base}_{2R}
+\vert N(\th)\rangle_1^{\rm fibre}\right)
\otimes\vert N(\tilde y_0)\rangle^{\rm base}_{2R},
\label{eqn:bulkDN}\nn\\
\vert N(\th)D(y_0)\rangle
&=&\frac{1+\sigma}{\sqrt 2}\vert N(\th)\rangle_1^{\rm fibre}\otimes
\vert D(y_0)\rangle^{\rm base}_{2R}\nn\\
&=&\frac{1}{\sqrt 2}\left(
\vert N(\th)\rangle_1^{\rm fibre}\otimes\vert D(y_0)\rangle^{\rm base}_{2R}
+\vert D(\th)\rangle_1^{\rm fibre}\otimes\vert D(y_0+2\pi R)\rangle^{\rm base}_{2R}
\right),
\label{eqn:bulkND}\nn\\
\vert N(\th)N(\tilde y_0)\rangle
&=&\frac{1+\sigma}{\sqrt 2}\vert N(\th)\rangle_1^{\rm fibre}\otimes
\vert N(\tilde y_0)\rangle^{\rm base}_{2R}\nn\\
&=&\frac{1}{\sqrt 2}\left(
\vert N(\th)\rangle_1^{\rm fibre}%\otimes\vert N(\tilde y_0)\rangle^{\rm base}_{2R}
+\vert D(\th)\rangle_1^{\rm fibre}\right)
\otimes\vert N(\tilde y_0)\rangle^{\rm base}_{2R}.
\label{eqn:bulkNN}
\eea
%\end{subequations}
%
It is obvious from the construction 
that these four states are actually not all distinct but only two are:
%%%%%%
\begin{eqnarray}
 && \ket{N(\th)D(y_0)}= \ket{D(\th)D(y_0+2\pi R)}, ~~~
\ket{N(\th)N(\ty_0)}= \ket{D(\th)N(\ty_0)}.
\end{eqnarray}
%%%%%%
%$\vert D(\th)D(y_0)\rangle$ and 
%$\vert N(\th)D(y_0)\rangle$
%represent the same class of states, so do
%$\vert D(\th)N(\tilde y_0)\rangle$ and
%$\vert N(\th)N(\tilde y_0)\rangle$.
%
It is straightforward to compute overlaps between these bulk brane states. 
Using the notation of (\ref{eqn:overlaps}) we find
($\Delta \th \equiv \th-\th'$,
$\Delta y_0 \equiv y_0-y_0'$, $\Delta \ty_0 \equiv \ty_0-\ty_0'$)
\bea
&&\langle D(\th)D(y_0)\vert e^{-\pi sH^{c}}\vert D(\th')D(y'_0)\rangle\nn\\
&&\hspace{30mm}
= Z_1^{DD}(it;\Delta \th)Z_{2R}^{DD}(it;\Delta y_0)
+Z^{DN}(it)Z_{2R}^{DD}(it;\Delta y_0+2\pi R),\\
&&\langle D(\th)N(\tilde y_0)\vert e^{-\pi sH^{c}}\vert D(\th')N(\tilde
y'_0)
\rangle\nn\\
&&\hspace{30mm}
= \left\{Z_{1}^{DD}(it;\Delta \th)+Z^{DN}(it)\right\}
Z_{2R}^{NN}(it;\Delta\tilde y_0),
%
%{}^{\rm bulk}\langle N(\tilde x_0)D(y_0)\vert e^{-\pi sH^{c}}\vert N(\tilde x'_0)D(y'_0)\rangle^{\rm bulk}
%&=& Z_{\sqrt{\alpha'}}^{NN}(\tau;\Delta\tilde x_0)Z_{2R}^{DD}(\tau;\Delta y_0)\\
%&&+Z^{DN}(\tau)Z_{2R}^{DD}(\tau;\Delta y_0+2\pi R),\\
%
%{}^{\rm bulk}\langle N(\tilde x_0)N(\tilde y_0)\vert e^{-\pi sH^{c}}\vert 
%N(\tilde x'_0)N(\tilde y'_0)\rangle^{\rm bulk}
%&=& \left\{Z_{\sqrt{\alpha'}}^{NN}(\tau;\Delta\tilde x_0)+Z^{DN}(\tau)\right\}
%Z_{2R}^{NN}(\tau;\Delta\tilde y_0),
\label{eqn:BBoverlaps1}
\eea
and 
\beq
\langle D(\th)D(y_0)\vert e^{-\pi sH^{c}}\vert D(\th')N(\tilde y'_0)
\rangle %\nn\\
%&&={}^{\rm bulk}\langle D(x_0)D(y_0)\vert e^{-\pi sH^{c}}\vert N(\tilde x'_0)N(\tilde y'_0)\rangle^{\rm bulk}
%\nn\\
%&&={}^{\rm bulk}\langle D(x_0)N(\tilde y_0)\vert e^{-\pi sH^{c}}\vert N(\tilde x'_0)D(y'_0)\rangle^{\rm bulk}
= \left\{Z_{1}^{DD}(it;\Delta \th)+Z^{DN}(it)\right\}Z^{DN}(it).%\nn\\
%
%&&{}^{\rm bulk}\langle D(x_0)D(y_0)\vert e^{-\pi sH^{c}}\vert N(\tilde x'_0)D(y'_0)\rangle^{\rm bulk}\nn\\
%&&= Z^{DN}(\tau)Z_{2R}^{DD}(\tau;\Delta y_0)
%+Z_{\sqrt{\alpha'}}^{DD}(\tau;\Delta x_0)Z_{2R}^{DD}(\tau;\Delta y_0+2\pi R),
%\nn\\
%
%&&{}^{\rm bulk}\langle D(x_0)N(\tilde y_0)\vert e^{-\pi sH^{c}}
%\vert N(\tilde x'_0)N(\tilde y'_0)\rangle^{\rm bulk}
%= \left\{Z_{\sqrt{\alpha'}}^{DD}(\tau;\Delta x_0)+Z^{DN}(\tau)\right\}Z_{2R}^{NN}(\tau;\Delta\tilde y_0),
%\nn\\
%
%&&{}^{\rm bulk}\langle N(\tilde x_0)D(y_0)\vert e^{-\pi sH^{c}}
%\vert N(\tilde x'_0)N(\tilde y'_0)\rangle^{\rm bulk}
%=\left\{Z_{\sqrt{\alpha'}}^{NN}(\tau;\Delta\tilde x_0)+Z^{DN}(\tau)\right\}Z^{DN}(\tau).
\label{eqn:BBoverlaps2}
\eeq
It is easy to 
check that each overlap represents a sum of Virasoro characters with non-negative integer multiplicity in the open string sector, satisfying the Cardy conditions.

%%%%%%%%%%%%%%%%%%%%%%%%%%%%%%%%%%%%%%%%%%%%%%%%%%%
%%%%%%%%%%%%%%%%%%%%%%%%%%%%%%%%%%%%%%%%%%%%%%%%%%%

%%%
\subsection{Fractional branes}
\label{sec:fracB}
%%%

One way of constructing fractional branes is to use the fact that the fibre CFT of the T-fold that we are considering is rational with respect to extended algebra ${\C A}_4\simeq{\C A}_1/{\B Z}_2$
(in the notation of \cite{Dijkgraaf:1989hb}; see Appendix \ref{sec:RCFT}).
Let us recall construction of Ishibashi states in rational conformal theory first. 
We assume the theory to be diagonal and look for boundary states that conserve the whole chiral algebra.
The conservation of the chiral symmetry on the boundary is characterised by trivial gluing conditions of the generators on the boundary states,
%\footnotemark[8],  
\beq
(W_m-(-1)^{h_W} \overline{W}_{-m})\vert B\rangle=0,
\label{eqn:Wconservation}
\eeq
where $W_m$ (${\overline W}_m$) are the mode operators of the left (right) chiral algebra generators,
and $h_W$ is the spin of the $W$ operator ($h_W=h_{\overline W}$).
The conditions (\ref{eqn:Wconservation}) include as a special case the
conformal invariance (Ishibashi) conditions,
\beq
(L_m-\overline{L}_{-m})\vert B\rangle =0,
\label{eqn:ishibashi}
\eeq
meaning that the left and right stress tensors are analytic on the boundary,
$[T-\bar T]_{\partial\Sigma}=0$. 
As the condition (\ref{eqn:Wconservation}) is linear any linear sum of $\vert B\rangle$ also satisfies this condition. 
A standard choice of basis in the space of such states is the Ishibashi states 
\beq
\vert \alpha\rish=\sum_N\vert \alpha;N\rangle\otimes U\overline{\vert \alpha;N\rangle},
\label{eqn:IshibashiStates}
\eeq
where $\alpha$ is the label for modules and $N$ is the label for states within each module. 
The anti-unitary operator $U$ comes from time reflection. 
An Ishibashi state (\ref{eqn:IshibashiStates}) intertwines the left and right Hilbert spaces; 
as the chiral blocks are irreducible representations of the chiral symmetry it follows from 
Schur's lemma that the intertwiners must be trivial.
We have seen in Sec.\ref{sec:TfoldCFT} that the T-fold with the self-dual fibre may be reformulated (using $T''$) so that the state space factorises into the left and right sectors that are isomorphic to each other. 
This allows us to use Ishibashi states of the form (\ref{eqn:IshibashiStates}) to analyse D-branes of the 
T-fold.

The left part of the fibre is a compact chiral boson on $S^1$ 
at radius $1/2$.
The theory is rational with respect to the extended symmetry ${\C A}_4$, 
with eight primary fields $\phi_{k=0,..,7}$.
They are realised by vertex operators
\beq
\phi_k(z)=e^{ikX_L(z)/2}.
\eeq
The right part of the fibre is a ${\B Z}_2$ orbifold chiral boson at self-dual orbifold radius. 
It is rational with respect to chiral algebra ${\C A}_1/{\B Z}_2$ and has eight primaries: 
the identity ${\B I}$, the current operator $j$, a pair of operators $\phi^i_2$ ($i=1,2$) that are inherited from the parent $S^1$ CFT, and four twist operators $\sigma^i$ and $\tau^i$. 
Basic features of these rational theories are summarised in Appendix \ref{sec:RCFT}.
The two chiral boson theories of the left and right parts of the fibre are equivalent, 
and there exists a one-to-one correspondence between the states. 
The correspondence of the rational CFT primaries is summarised in Table 1.

%%%%%
%Table 1
%%%%%
\begin{center}
\begin{tabular}{c||ccc|cc}
%\hline
% after \\ : \hline or \cline{col1-col2} \cline{col3-col4} ...
Sector &\multicolumn{3}{c|}{T-even (untwisted)}&\multicolumn{2}{c}{T-odd (twisted)}\\
\hline
Conformal weight & $0$ & $1$ & $\quarter$ & $\frac{1}{16}$ &$\frac{9}{16}$\\
\hline
${\C A}_4$ primary 
& $\phi_0$ & $\phi_4$ & $\phi_2$, $\phi_6$ & $\phi_1$, $\phi_7$ & $\phi_3$, $\phi_5$ \\
${\C A}_1/{\B Z}_2$ primary 
& ${\B I}$ & $j$ & $\phi^1_2$, $\phi^2_2$ & $\sigma^1$, $\sigma^2$ & $\tau^1$, $\tau^2$
%\hline
\end{tabular}\\
\vspace{10pt}
{\bf Table 1.} Correspondence of the primary fields in ${\C A}_4$ and ${\C A}_1/{\B Z}_2$
rational theories.
\end{center}
%%%
%
%
As being the same chiral CFT the correspondence is not limited to the level of rational CFT primaries 
but persists also at the level of the Virasoro primaries.
It is convenient to introduce an isomorphic map $\iota$ from a state of the ${\C A}_4$ CFT to the corresponding state in the ${\C A}_1/{\B Z}_2$ CFT. 
Using this map we may write, for example, $\iota\vert\phi_0\rangle=\vert{\B I}\rangle$.
Eight Ishibashi states corresponding to the eight rational primaries of the fibre of the T-fold are then,
\beq
\vert \phi_k\rish=\sum_N\vert \phi_k;N\rangle\otimes \overline{\vert\iota\phi_k;N\rangle}.
\label{eqn:T-Ishibashi}
\eeq
Note that $\phi_k$ are the ${\C A}_4$ primary labels, while $\iota\phi_k$ refer to 
${\C A}_1/{\B Z}_2$ primaries.
The Cardy boundary states are found in the usual way (see (\ref{eqn:AnCardy}) below),
\beq
|\phi_k\rangle_C=2^{-\frac 34}\sum_{\ell=0}^7e^{-i\pi k\ell/4}\vert\phi_\ell\rish,
\label{eqn:phikC}
\eeq
using the fibre Ishibashi states defined above.
These are linear sums of T-even states $\vert\phi_{0,2,4,6}\rish$ and T-odd states 
$\vert\phi_{1,3,5,7}\rish$.
We have to choose Neumann condition on the base as it is invariant under the shift
${\C T}_{2\pi R}: Y\rightarrow Y+2\pi R$. 
With Wilson line $\tilde y_0\in S^1$ turned on, the base Neumann state is
\beq
\vert N(\tilde y_0,\alpha)\rangle^{\rm base}
=2^{1/4}\sqrt{R} \sum_{w\in 2{\B Z}+\alpha}e^{-iw\tilde y_0 R} 
e^{-\sum_{m=1}^\infty \frac 1m b_{-m}\bar b_{-m}}\vert (n=0,w)\rangle^{\rm base},
\label{eqn:BaseNeumannIshibashi}
\eeq
where $b_m$, $\bar b_m$ are the left and right mode operators of the 
base field $Y$, and $\alpha=0$ ($1$) in the untwisted (twisted) sector.
Fractional boundary states of the full T-fold theory is found 
by combining the fibre with the base in such a way that the fibre is T-even 
(odd) when the base winding number is even (odd).
This is analogous to the construction of the one-loop partition function.
We thus find fractional brane states,
\bea
\vert \phi_k^{\rm fibre} N^{\rm base}(\tilde y_0) \rangle
&=& \sqrt{\frac{R}{2}}  \left\{
\sum_{\stackrel{\scriptstyle w\in 2{\B Z}}{\ell=0,2,4,6}}
e^{-iw\tilde y_0 R-\frac{i\pi k\ell}{4}}
e^{-\sum_{m=1}^\infty \frac{b_{-m}\bar b_{-m}}{m}}
\vert\phi_\ell\rish^{\rm fibre}\vert (0,w)\rangle^{\rm base}\right.\nn\\
&&+\left.\sum_{\stackrel{\scriptstyle w\in 2{\B Z}+1}{\ell=1,3,5,7}}
e^{-iw\tilde y_0 R-\frac{i\pi k\ell}{4}}
e^{-\sum_{m=1}^\infty\frac{b_{-m}\bar b_{-m}}{m}}
\vert\phi_\ell\rish^{\rm fibre}
\vert(0,w)\rangle^{\rm base}\right\}.
\label{eqn:phik_N}
\eea
The fibre is characterized by the RCFT primary of 
$({\C A}_4)^L\otimes({\C A}_1/{\B Z}_2)^R$ 
and the label is taken from the left part ($\phi_k$). 
It is not quite correct to call the fibre part as Neumann or Dirichlet;
the Cardy states of the ${\C A}_1/{\B Z}_2$ CFT (the right-moving sector) are 
identified with 4 Dirichlet and 4 Neumann states at the orbifold fixed points, 
while those of the ${\C A}_4$ theory (the left-moving sector) may be identified 
as Neumann states with Wilson line values at evenly spaced 8 points on the 
$S^1$, that is $\tilde x_0=0, \half\pi, \pi, \cdots, \frac 72\pi$.
Although there is no naturally defined momentum 
or winding number in the twisted sector of the fibre,
it is clear from the construction that one may introduce 
ground states $\vert [n,w]\rangle$ 
with the momentum $n$ and the winding 
$w$ inherited from the left-moving ${\C A}_4$ CFT.
Introducing also mode operators $\bar a'_m$ ($m\in{\B Z}$) in the
right-moving sector that correspond to $a_n$ in the left-moving sector,
one may identify the Cardy states (\ref{eqn:phikC}) with `Neumann'
states having Wilson line 
$\tilde x_0=\frac{k\pi}{2}$, 
\beq
\vert\phi_k\rangle_C
=2^{-\frac 34}\sum_{w\in{\B Z}}e^{-\frac{i\pi wk}{4}} 
e^{-\sum_{m=1}^\infty\frac 1n a_{-m}\bar a'_{-m}}\vert [0,w]\rangle^{\rm fibre}.
\eeq
In this notation the Ishibashi states may be written as,
\beq
\vert\phi_\ell\rish=
e^{-\sum_{m=1}^\infty\frac 1m a_{-m}\bar a'_{-m}}\sum_{w\in{\B Z}}
\vert [0,\ell+8w]\rangle^{\rm fibre}.
\label{eqn:Ishi}
\eeq
Inserting (\ref{eqn:Ishi}) into (\ref{eqn:phik_N}) one may rewrite the fractional states as
\bea
&&\vert \phi_k^{\rm fibre} N^{\rm base}(\tilde y_0)\rangle \nn\\
&&=\sqrt{\frac{R}{2}}
\left\{
\sum_{\stackrel{\scriptstyle w\in 2{\B Z}}{\ell=0,2,4,6}}
e^{-iw\tilde y_0 R-\frac{i\pi k\ell}{4}}
e^{-\sum_{m=1}^\infty\frac 1m(a_{-m}\bar a'_{-m}+b_{-m}\bar b_{-m})}
\sum_{w'\in{\B Z}}\vert [0,\ell+8w']\rangle^{\rm fibre}\vert(0,w)\rangle^{\rm base}\right.\nn\\
&&+\left.\sum_{\stackrel{\scriptstyle w\in 2{\B Z}+1}{\ell=1,3,5,7}}
e^{-iw\tilde y_0 R-\frac{i\pi k\ell}{4}}
e^{-\sum_{m=1}^\infty\frac 1m (a_{-m}\bar a'_{-m}+b_{-m}\bar b_{-m})}
\sum_{w'\in{\B Z}}\vert [0,\ell+8w']\rangle^{\rm fibre}\vert(0,w)\rangle^{\rm base}\right\}.
\eea
Recalling that $k$ is related to the value of the Wilson line of the ${\C A}_4$ theory by
$\half k\pi=\tilde x_0\equiv\theta$ one may write the fractional states parametrised by 
$\theta$ and $\tilde y_0$:
\beq
\vert F; \theta, \tilde y_0\rangle
%&=&\sqrt{\frac{R}{2}}
%\left\{
%\sum_{w,\ell\in2{\B Z}}
%e^{-iw\tilde y_0 R-\frac{i\ell \theta}{2}}
%e^{-\sum_{m=1}^\infty \frac 1m(a_{-m}\bar a'_{-m}+b_{-m}\bar b_{-m})}
%\vert [0,\ell]\rangle^{\rm fibre}\vert(0,w)\rangle^{\rm base}\right.\nn\\
%&&+\left.\sum_{w,\ell\in2{\B Z}+1}
%e^{-iw\tilde y_0 R-\frac{i\ell \theta}{2}}
%e^{-\sum_{m=1}^\infty \frac 1m(a_{-m}\bar a'_{-m}+b_{-m}\bar b_{-m})}
%\vert [0,\ell]\rangle^{\rm fibre}\vert(0,w)\rangle^{\rm base}\right\}\nn\\
= \sqrt{\frac{R}{2}}
\sum_{\stackrel{\scriptstyle w,\ell\in {\B Z}}{w-\ell\in2{\B Z}}}
e^{-iw\tilde y_0 R-\frac{i\ell \theta}{2}}
e^{-\sum_{m=1}^\infty \frac 1m(a_{-m}\bar a'_{-m}+b_{-m}\bar b_{-m})}
\vert [0,\ell]\rangle^{\rm fibre}\vert(0,w)\rangle^{\rm base}.
\label{eqn:FFF}
\eeq
Here the parameter $\theta$ may be regarded continuous, reflecting unbroken 
$U(1)$ symmetry of the moduli. 
It is periodic and we take its range as $0\leq \theta<4\pi$.

%%%%%%%%%%%%%%%%%%%%%%%%%%%%%%%%%%%%%%%%%%%%%%%%%%%

Instead of the somewhat cluttered bottom up approach 
described above one may formulate the fractional states 
by focusing on the underlying $SU(2)_1$ symmetry of the fibre.
An advantage of this method is that it is easier to evaluate cylinder amplitudes
with the bulk branes.
We start with recalling that the T-duality operator $T''$ (\ref{eqn:TonJs}) 
acts as asymmetric rotations on the fibre
\beq
T'' = (e^{i\pi J^3_0}, e^{i\pi \bar J^1_0}),
\label{eqn:T-action3}
\eeq
with $J^a$ and $\bar J^a$ the left and right $SU(2)_1$ currents.
It is convenient to introduce an automorphism $\kappa$ of $SU(2)_1$, 
defined by 
\begin{eqnarray}
 && \kappa J^1 \kappa^{-1} = J^3, ~~~
  \kappa J^2 \kappa^{-1}= J^2, ~~~ 
  \kappa J^3 \kappa^{-1}= - J^1, ~~~ 
  \kappa \bar J^a \kappa^{-1}= \bar J^a.
\label{kappa}
\end{eqnarray}
The point is that $\kappa$ interpolates 
between the standard reflection orbifold and the orbifold generated by $T''$.
Indeed, the reflection of 
the fibre ${\C R}\,:\,X=(X_L, X_R)\,\rightarrow\,-X$ may be written
\beq
{\C R} = (e^{i\pi J^1_0}, e^{i\pi \bar J^1_0})~,
\eeq
and hence
\beq
\kappa {\C R} \kappa^{-1} = T''.
\eeq
Note that for the untwisted Hilbert space (
i.e. the integrable reps. of $SU(2)_1$), 
we may explicitly write 
\begin{eqnarray}
 && \kappa = e^{i\frac{\pi}{2}J^2_0}~, ~~~
  \kappa^{-1}= e^{-i\frac{\pi}{2}J^2_0}~.
\label{eqn:kappainJ2}  
\end{eqnarray}
In the twisted sector $\kappa$ cannot be written as (\ref{eqn:kappainJ2}) since
$J^2$ does not have zero-mode on ${\C H}^{{\C R}}$ or ${\C H}^{T''}$.
It is nevertheless clear\footnote
{
Recall the discussion based on the rational CFT primaries. 
One may identify $\kappa$ as 
$$
\kappa = (\iota^{-1}, {\bf 1})~: ~ (\cA_1/{\B Z}_2)^L\otimes (\cA_1/{\B Z}_2)^R
~\stackrel{\cong}{\longrightarrow}~
\cA_4^L \otimes (\cA_1/{\B Z}_2)^R.
$$
}
that $\kappa$ may be extended to isomorphism 
between the ${\C R}$-twisted Hilbert space ${\C H}^{{\C R}}$ 
and the $T''$-twisted Hilbert space ${\C H}^{T''}$,
\beq
\kappa~:~{\C H}^{{\C R}}~\stackrel{\cong}{\longrightarrow}~
{\C H}^{T''}.
\eeq
We wish to find boundary conditions that are invariant under operation of 
$\sigma \equiv T'' \otimes {\C T}_{2\pi R}$.
Along the base circle we have to 
choose as before Neumann conditions as they are invariant under the shift, 
${\C T}_a \vert N;\tilde y_0\rangle_{2R}= \vert N;\tilde y_0\rangle_{2R}$.
On the fibre desired boundary states are obtained from the usual reflection orbifold by using the interpolation $\kappa$.
%%%%%%%%%%%%%%%%%%%%%%%%
%%%%%%%%%%%%%%%%%%%%%%%%
There is a one-parameter (Wilson line) 
family of fractional branes in the reflection
orbifold
\footnote{
	These are periodic in $\theta$ with periodicity $4\pi$,   
	$\vert{F;\theta + 4\pi}\rangle^{{\C R}} = \vert{F;\theta}\rangle^{{\C R}}$.
	Note that $\vert{F;\theta+2\pi}\rangle^{{\C R}}
	= e^{2i\theta J^1_0} (\vert{N}\rangle_{1}-\vert{N}\rangle_{1}^{{\C R}})/ \sqrt 2$. 
	They represent Neumann condition when $\theta=n \pi$ and Dirichlet condition when 
	$\theta= \frac{\pi}{2}+ n\pi$ ($n=0,\ldots,3$). 
}:
\beq
 \vert{F;\theta}\rangle^{{\C R}} = 
 \frac{e^{2i\theta J^1_0}}{\sqrt 2} \left(\vert{N}\rangle_1
 +\vert{N}\rangle_1^{{\C R}}\right),
\label{eqn:RefOrbState} 
\eeq
which is obviously reflection invariant, 
${\C R} \vert{F;\theta}\rangle^{{\C R}} = \vert{F;\theta}\rangle^{{\C R}}$. 
The Neumann boundary state in the ${\C R}$-twisted 
sector $\vert{N}\rangle_1^{{\C R}}$ is 
characterized by
\begin{eqnarray}
&&(J^1_n+\bar J^1_{-n})\vert{N}\rangle_1^{{\C R}}=0, ~~~ 
(n\in {\B Z})~, \nn\\
&&(J^a_r+\bar J^a_{-r})\vert{N}\rangle_1^{{\C R}}=0, ~~~ 
(r\in \frac{1}{2}+{\B Z}, ~~~ a = 2,3)~,  \nn\\
&&{}_{1}^{{\C R}}\langle{N}\vert e^{-\pi s H^{(c)}}
  e^{2\pi i z J^1_0} \vert{N}\rangle^{{\C R}}_1 
= \frac{\Theta_{1/2,1}(z|is) + \Theta_{-1/2,1}(z|is)} {\sqrt{2}\eta(is)} \nn\\
&&\hspace{40mm}=\frac{1}{\eta(it)}\sum_{n\in{\B Z}} 
(-1)^n e^{-2\pi t \left(n+\frac{z}{2}\right)^2}. \hspace{16mm} (t\equiv 1/s)
\label{eqn:RtwistedN}
\end{eqnarray}
Fractional branes of the $T''$-orbifold are obtained from (\ref{eqn:RefOrbState}) using $\kappa$,
\beq
\vert{F;\theta}\rangle^{T''} 
= \kappa \vert{F;\theta}\rangle^{{\C R}}
= \kappa\frac{e^{2i\theta J^1_0}}{\sqrt 2} \left(\vert{N}\rangle_1
+\vert{N}\rangle_1^{{\C R}}\right)
= \frac{e^{2i\theta J^3_0}}{\sqrt 2} \kappa  \left(\vert{N}\rangle_1
+\vert{N}\rangle_1^{{\C R}}\right).
\label{eqn:T''OrbState}
\eeq
Fractional brane states of the T-fold model associated with the combined operation 
$\sigma = T'' \otimes {\C T}_{2\pi R}$ then read 
\begin{eqnarray}
 \vert{F;\theta, \tilde y_0}\rangle 
% &=& \vert{N;\tilde y_0}\rangle_{2R} \otimes 
%\frac{1+T''}{2\sqrt{2}} e^{2i\theta J^3_0} \kappa \vert{N}\rangle_{1}
%+ \vert{N;\tilde y_0}\rangle_{2R}^{\C T} \otimes 
% \frac{1+T''}{2\sqrt{2}} e^{2i\theta J^3_0} \kappa \vert{N}
%\rangle^{{\C R}}_{1}\nn\\
&=& \frac{1}{\sqrt{2}}
\vert{N (\tilde y_0)}\rangle_{2R} 
\otimes e^{2i\theta J^3_0} \kappa \vert{N}\rangle_{1}
+ \frac{1}{\sqrt{2}}\vert{N (\tilde y_0)}\rangle_{2R}^{\C T} \otimes 
 e^{2i\theta J^3_0} \kappa \vert{N}\rangle^{{\C R}}_{1},
\label{eqn:FracBrane1}
\end{eqnarray} 
which represents the same states as those found earlier (\ref{eqn:FFF}).
These states are invariant under the orbifold projection, 
$\half (1+\sigma)\vert F;\theta,\tilde y_0\rangle=\vert F;\theta,\tilde y_0\rangle$.
The Neumann states of the base circle in the twisted sector
$\vert{N (\tilde y_0)}\rangle_{2R}^{\C T}=\vert N (\tilde y_0, \alpha=1)\rangle^{\rm base}$ 
(see  (\ref{eqn:BaseNeumannIshibashi}))
are characterized by (with the same normalisation as 
$\vert{N (\tilde y_0)}\rangle_{2R}=
\vert{N (\tilde y_0, \alpha=0)}\rangle^{\rm base}$ 
in the untwisted sector)
\begin{eqnarray}
	{}^{\C T}_{2R}\langle{N(\tilde y_0)}\vert e^{-\pi s H^{(c)}}
	\vert{N (\tilde y'_0)}\rangle^{\C T}_{2R}
	& =& 
\frac{2R}{\sqrt{2}}\frac{1}{\eta(is)}\sum_{w\in {\B Z}} e^{-2\pi s \frac{1}{4}
	\left\{2R\left(w+\frac{1}{2}\right)\right\}^2}
e^{-i2R\left(w+\frac{1}{2}\right)(\Delta\tilde y_0)}\nn\\
	&=& \frac{1}{\eta(it)} \sum_{n\in {\B Z}} (-1)^n e^{-2\pi t 
	\left(\frac{n}{2R}+\frac{\Delta\tilde y_0}{2\pi}\right)^2} .
\end{eqnarray}

%%%%%%%%%%%%%%%%%%%%%%%%%%%%%%%%%%%%%%%%%%%%%%%%%%%
Let us evaluate the overlaps involving the fractional states.
We first consider the overlaps with the bulk brane states.
Clearly, only the untwisted sector contributes to the amplitudes, 
and we may simply replace $\kappa$ with $e^{i\frac{\pi}{2}J^2_0}$. 
We can then utilize the $SU(2)_1$ technique demonstrated in Appendix 
\ref{sec:MixedFormulae}. 
%%%%%%%%%%%%%%%%%%%%%%%%%%%%%%%%%%%%%%%%%%%%%%%%%%%
Making use of (\ref{eqn:formula1}) we find ($\Delta \th \equiv
\th-\th'$,
$\Delta \ty_0 \equiv \ty_0 -\ty_0'$),
\begin{eqnarray}
% && \langle{DN;\tilde y_1,\tilde x_1}\vert e^{-\pi s H^{(c)}} \vert{F;\tilde y_2,\theta_2}\rangle= 
\langle{D(\theta)D(y_0)}\vert e^{-\pi s H^{(c)}} 
\vert{F; \theta', \tilde y_0'}\rangle %\nn\\
%&& \hspace{2cm}
%&=& 
%\langle{N(\theta)D(y_0)}\vert e^{-\pi s H^{(c)}} 
%\vert{F; \theta', \tilde y_0'}\rangle \nn\\
&=&  Z^{DN}(it) \frac{1}{\eta(it)}
\sum_{n\in {\B Z}} 
e^{-2\pi t \left\{n+\frac{1}{2\pi}\alpha(\Delta \th)\right\}^2},\nn\\
%%%
% && \langle{NN;\tilde y_1,\tilde x_1}\vert e^{-\pi s H^{(c)}} \vert{F;\tilde y_2,\theta_2}\rangle= 
\langle{D(\th)N(\tilde y_0)}\vert e^{-\pi s H^{(c)}}\vert{F; \theta',
\tilde y'_0}\rangle %\nn\\
%&& \hspace{2cm} 
%&=& 
%\langle{N(\th)N(\tilde y_0)}\vert e^{-\pi s H^{(c)}}\vert{F; \theta',
%\tilde y'_0}\rangle \nn\\
&=& Z^{NN}_{2R}(it;\Delta\tilde y_0) \frac{1}{\eta(it)} 
\sum_{n\in {\B Z}} e^{-2\pi t \left\{n+ \frac{1}{2\pi}\alpha(\Delta \th)\right\}^2},
\label{eqn:BFoverlaps}
\end{eqnarray}
where we introduced the notation 
\begin{eqnarray}
 \alpha(z) = \cos^{-1}
\left(\frac{\cos z}{\sqrt{2}}\right).
\end{eqnarray}

In computing the overlaps between the fractional branes one can evaluate the untwisted and twisted pieces separately. 
In the untwisted sector we find,
\beq
\langle{F; \theta, \tilde y_0}\vert e^{-\pi s H^{(c)}} 
\vert{F; \theta', \tilde y'_0}\rangle \left|_{\msc{untwisted}} \right.
= \half Z^{NN}_{2R}(it; \Delta\tilde y_0) Z^{NN}_1(it; \Delta\theta)~,
\eeq
and in the twisted sector,
\beq
\langle{F; \theta, \tilde y_0}\vert e^{-\pi s H^{(c)}} 
\vert{F; \theta', \tilde y'_0}\rangle\left|_{\msc{twisted}} \right.
= \frac{1}{2\eta(it)^2} \sum_{m,n\in {\B Z}} (-1)^{m+n}
e^{-2\pi t \left\lb(\frac{m}{2R}+ \frac{\Delta\tilde y_0}{2\pi})^2+(n+ \frac{\Delta\theta}{2\pi})^2\right\rb}~.
\eeq
The total fractional-fractional overlap is then,
\beq
\langle{F; \theta, \tilde y_0}\vert e^{-\pi s H^{(c)}} 
\vert{F; \theta', \tilde y'_0}\rangle
=\frac{1}{\eta(it)^2}\sum_{\stackrel{\scriptstyle m,n\in{\B
Z}}{m-n\in 2{\B Z}}}
e^{-2\pi t \left\lb(\frac{m}{2R}+\frac{\Delta\tilde
y_0}{2\pi})^2+(n+\frac{\Delta\theta}{2\pi})^2\right\rb}.
\label{eqn:FFoverlap}
\eeq
The amplitudes (\ref{eqn:BFoverlaps}) and (\ref{eqn:FFoverlap}) display 
$q$-expansions ($q\equiv e^{-2\pi t}$) 
with non-negative integer multiplicities in the open string channel,
and hence the Cardy conditions are satisfied. 
When the boundary conditions on the two boundaries are same the amplitude (\ref{eqn:FFoverlap}) contains the Virasoro vacuum character with multiplicity one, indicating that the boundary states
(\ref{eqn:FracBrane1}) represent elementary fractional branes.  

%%%
\subsection{Some comments on the branes}
%%%

%%%%%%%%%%%%%%%%%%%%%%%%%%%%%%%%%%%%%%%%%%%%%%%%%%%
We conclude this section with comments on the D-branes we have found.
%%%%%%%%%%%%%%%%%%%%%%%%%%%%%%%%%%%%%%%%%%%%%%%%%%%

\noindent
{\bf 1. } The bulk branes allow obvious geometrical interpretations. 
The branes localized on the base circle (Dirichlet b.c. along the base) are
interpretable as an alternating array of D0 and D1 branes along the
fiber if lifted up to the universal cover of the base circle. Also, a
brane wrapped on the base (Neumann b.c. along the base) 
is nothing but a superposition of D0 and D1 branes along the fiber
which are T-dual to each other. 
These branes should be consistent with those given by the classical 
analysis based on the doubled torus approach \cite{Lawrence:2006ma}. 
This is obvious for the branes localized on base. 
It is also inferred by arguments in \cite{Lawrence:2006ma} 
that the consistent branes wrapped on the base must have even winding
numbers. 
This in fact agrees with our analysis as the bulk boundary states with Neumann b.c. 
along the base (the $DN$ and $NN$ states in (\ref{eqn:bulkNN})) are identified with 
branes wrapped twice on the base; those wrapped only once cannot exist consistently
as a geometric object in the doubled torus.

%%%%%%%%%%%%%%%%%%%%%%%%%%%%%%%%%%%%%%%%%%%%%%%%%%%

~

\noindent
{\bf 2. } The fractional branes on the other hand are more curious
as they do not have a simple geometrical interpretation. 
One can for example read from the cylinder amplitudes \eqn{eqn:BFoverlaps} 
that the lightest mass of an open string between a fractional and a bulk brane
is a non-linear function of the moduli of the branes (location or Wilson line along the fiber),
$\propto \left\lb \cos^{-1} \left(\frac{\cos z}{\sqrt{2}}\right)\right\rb$, with $z$ the modulus. 
%Moreover, such an open string can never be massless.
This feature appears to be rather 
exotic compared to the standard D-brane dynamics
on geometric backgrounds.
We point out that the physics of T-fold may be distinguished by
this characteristic feature from a geometric background
(i.e. a non-linear $\sigma$-model),
even at energy scales much lower than the string scale.
This is due to the non-linear behavior mentioned above already appearing 
{\em in the no-winding sector of the base circle}. 
On the other hand, if looking at the closed string sector, 
non-geometric properties of T-fold originate only from strings wound 
(odd times) around the base circle, which are expected to 
decouple from the low energy physics. 
For this reason D-brane dynamics would be important in investigating 
physics of T-folds. 
%%%%

%%%%%%%%%%%%%%%%%%%%%%%%%%%%%%%%%%%%%%
%%%%%%%%%%%%%%%%%%%%%%%%%%%%%%%%%%%%%%

Let us be more specific about what we actually mean by `geometric' or
`non-geometric.'
We classify the boundary conditions defining D-branes into two classes:
(i) `geometric branes,' corresponding to linear gluing conditions with 
respect to the $\sigma$-model coordinates $X$, $Y$, and 
(ii) `non-geometric branes,' defined by non-linear gluing conditions
\footnote
   {If one instead describes the $X$-sector by the 
   $SU(2)$-WZW model at level 1, all the boundary conditions 
   considered here are linearly realized in terms of the 
   $SU(2)$-current algebra. 
   Note, however, the $SU(2)$-WZW model is quite different from a non-linear 
   $\sigma$-model which has the central charge equal to the dimensionality 
   of the target space.  }. 
Geometric branes in this sense have obvious interpretations in terms of 
non-linear $\sigma$-models with boundaries, 
whereas non-geometric branes are not. 
Geometric branes are of primary importance as objects in the classical geometry  
defined in the particle theory limit. 
The bulk branes considered above are actually geometric in this sense.
On the other hand, in the reflection orbifold 
$S^1_{R'=1}/{\B Z}_2\cong 
\left(\cA_1/{\B Z}_2\right)^L\otimes \left(\cA_1/{\B Z}_2\right)^R$, 
the fractional brane \eqn{eqn:RefOrbState} 
has one modulus parameter $\theta$, 
and there exist eight geometric points corresponding to linear boundary conditions
in the moduli space:
$\theta=n\pi$ (Neumann), $\theta=\frac{\pi}{2}+n\pi$ (Dirichlet)
with $n=0,1,2,3$ (see also Appendix \ref{sec:RCFT}.)
In contrast, our fractional branes in the T-fold are entirely
non-geometric since the boundary condition is always non-linear
in the moduli space.

%%%%%%%%%%%%%%%%%%%%%%%%%%%%%%%%%%%%%%%%%%%%%%%%%%%
We emphasise that, if comparing the T-fold with 
the symmetric orbifold (reflection orbifold), 
the spectra of Cardy states 
with respect to Virasoro algebra should 
be identical, since the torus partition functions
coincide and thus they have isomorphic Hilbert spaces
of closed string states. What we address here is that 
they nevertheless have inequivalent spectra 
of {\em geometric branes}. 
The geometric bulk branes in the T-fold we constructed above
are mapped by the isomorphism to some 
non-geometric branes in the reflection
orbifold, and {\em vice versa}.  
Moreover, as is obvious from our construction, 
the fractional branes in the T-fold are mapped to those  
in the reflection orbifold by the isomorphism;
the latter are well-defined geometrical objects localized at the fixed points 
of the orbifold (and their marginal boundary deformations), whereas the
former are entirely non-geometric, as addressed above. 
%%%%%%%%%%%%%%%%%%%%%%%%%%%%%%%%%%%%%%%%%%%%%%%%%%%%%%%%%

It is not clear to us at the moment 
how the fractional branes 
may be understood in the framework of the doubled torus. 
This is obviously an interesting issue.
It might be of some help to consider the model as a special case of the $SU(2)$ WZW model 
(see section 5).
%%%%%

%%%%%%%%%%%%%%%%%%%%%%%%%%%%%%%%%%%%%%%%%%%%%%%%%%%

~

\noindent
{\bf 3. }
% g-factor 
An important set of information encoded in the boundary states is the ground state degeneracy
(Affleck-Ludwig $g$-factor)\cite{Affleck:1991tk}.
It is defined as the overlap of a boundary state with the M\"obius -invariant untwisted closed string vacuum, 
\beq
g_B=\langle(n=0, w=0)\vert B\rangle,
\eeq
where the phase convention of the states are chosen so that $g_B\geq 0$.
The $g$-factor is a conformal fixed point value of the $g$-function that decreases along boundary renormalisation group flows (analogous to the celebrated $c$-theorem in the bulk). 
For $c=1$ CFT on $S^1$ of radius $R$, the $g$-factors of the Dirichlet and Neumann states are
\beq
g_D(R)=\frac{1}{2^{1/4}\sqrt{R}},\;\;\;
g_N(R)=\frac{\sqrt{R}}{2^{1/4}}.
\label{eqn:unorbg}
\eeq
When CFT under study appears as an internal space of string compactification (such as in our case), 
the $g$-factor measures the mass (or stability) of the brane\cite{Harvey:1999gq}.
The rationale behind this is that the mass of a brane is actually measured by its interaction with 
gravitons. 
The scattering amplitude is computed from the two-point function of graviton vertices on the disk topology, which reduces through bulk operator product expansions
to (a sum of) one point functions on the disk,
\beq
A^{\mu\nu}=\langle \vec{k}_L, \vec{k}_R \vert a^\mu_1\bar a^\nu_1\vert B\rangle,\;\;\; \mu,\nu=0,\ldots, D-1
\eeq
($D$ is the spacetime dimensions). 
Its symmetric traceless part yields the metric, the antisymmetric traceless part the Kalb-Ramond 2-form field, and the trace part the dilaton upon Fourier transformation\cite{DiVecchia:1997pr}. 
The $A^{\mu\nu}$ factorises into a noncompact spacetime part and a compact internal part. 
From the noncompact spacetime viewpoint the $g$-factor from the internal CFT appears universally as a coefficient of the graviton amplitude and contributes to the coupling strength of the graviton to the brane. 
The $g$-factor of D-branes in the T-fold is immediately read off from the boundary states.
They are
\beq
g^{D^{\rm fibre}D^{\rm base}}_{\rm bulk}(R)=%g^{N^{f}D^{b}}_{\rm bulk}=
\frac{1}{\sqrt{2R}},\;\;\;
g^{D^{\rm fibre}N^{\rm bulk}}_{\rm bulk}(R)=%g^{N^{f}N ^{b}}_{\rm bulk}=
\sqrt{2R},
\eeq
for the bulk brane states and
\beq
g_{\rm frac}(R)=\sqrt{\frac{R}{2}}
\eeq
for the fractional states.
%These may be compared with the $g$-factors (\ref{eqn:unorbg}) for unorbifolded 
%$S^1_{\sqrt{\alpha'}}\otimes S^1_{R}$.
%One finds,
%$g^{D^{\rm fibre}D^{\rm base}}_{\rm bulk}(R)=g_{sd}g_D(R)$, 
%$g^{D^{\rm fibre}N^{\rm base}}(R)=2g_{sd}g_N(R)$ and 
%$g_{\rm frac}(R)=g_{sd}g_N(R)$, where  
%$g_{sd}=g_D(\sqrt{\alpha'})=g_N(\sqrt{\alpha'})$.
As $g^{DD}_{\rm bulk}\ll g_{\rm frac}<g^{DN}_{\rm bulk}$ when $R\gg 1$ and
$g_{\rm frac}<g^{DN}_{\rm bulk}\ll g^{DD}_{\rm bulk}$ when $R\ll 1$,
we find from the above reasoning that the bulk branes with the Dirichlet base are most stable 
in the former case, whereas in the latter the fractional branes are most stable.

~

%%%%%%%%%%%%%%%%%%%%%%%%%%%%%%%%%%%%%%%%%%%%%%%%%%%

\noindent
{\bf 4. } It is also easy to construct boundary states in the
T-dualized T-fold \eqn{T-dualized T-fold}. All we have to do is to
exchange the Neumann and Dirichlet boundary states in the base part 
in \eqn{eqn:bulkDD}, \eqn{eqn:bulkDN}, \eqn{eqn:FracBrane1} etc. 
Especially, only the Dirichlet b.c. along the base direction
is possible for the fractional branes, since the `double cover' operator
$\widetilde{\cT}_{2\pi \frac{1}{R}}$ leaves the Dirichlet
b.c. invariant, while for the Neumann b.c. it does not.

%%%%%%%%%%%%%%%%%%%%%%%%%%%%%%%%%%%%%%%%%%%%%%%%%%%
%%%%%%%%%%%%%%%%%%%%%%%%%%%%%%%%%%%%%%%%%%%%%%%%%%%

%%%%%
\section{World-sheet fermions}
%%%%%

Our discussion so far has been limited to the bosonic theory.
We now consider a simple ${\C N}=1$ 
extension of the $S^1$ over $S^1$ T-fold that 
we have discussed in the previous sections. 
%Note however that this ${\C N}=1$ system cannot be embedded directly in type II superstring theory, as the T-duality exchanges IIA and IIB.
In addition to the fibre and base bosons $X$ and $Y$, we introduce the fibre and base fermions which we shall denote $\psi^X$ and $\psi^Y$.
Under the usual T-duality the fibre fields undergo transformations,
\beq
(X_L, X_R)\rightarrow (X_L, -X_R),\;\;\;
(\psi^X_L, \psi^X_R)\rightarrow (\psi^X_L, -\psi^X_R). 
\label{eqn:superT1}
\eeq
%suggesting that in the twisted sector the R and NS boundary conditions of the fibre are interchanged.
As it turns out, construction of a modular invariant partition function 
(while keeping the natural order 2 orbifold structure) is not entirely automatic.
Below we describe a model of ${\C N}=1$ T-fold that is an asymmetric orbifold of order 2; 
this is based on an observation that at a special radius of the fibre there exists a 
global $SU(2)$ symmetry which is similar to the one we encountered in the bosonic case.

We choose the fibre radius to be the free fermion radius $R=\sqrt{2}$ (the $SO(2)$-point).
This allows one to fermionise the fibre boson $X=X_L+X_R$ according to the rule
\beq
\frac{1}{\sqrt 2}(\psi_L^1\pm i\psi_L^2)=e^{\pm iX_L\sqrt{2}},
\eeq
and likewise for the right mover.
Identifying the fermionic component as 
\beq
\psi_L^3=\psi^X_L,
\eeq
the fibre is represented by a system of three fermions, 
which is known to possess an $SO(3)_1\cong SU(2)_2$ current algebra symmetry.
Indeed, the affine $SU(2)$ currents at level 2 are explicitly constructed as  
\beq
J^a=-i\epsilon^{abc}\psi_L^b\psi_L^c,~~~\tJ^a=-i\epsilon^{abc}\psi_R^b\psi_R^c, 
\label{eqn:su(2)at2}
\eeq
where $\epsilon^{abc}$ being totally antisymmetric and
$\epsilon^{123}=+1$.

%%%%%%%%%%%%%%%%%%%%%%%%%%%%%%%%%%%%%%%%%%%%%%%%%%%

We start with the diagonal spin structures and make an orbifolding
\begin{eqnarray}
 S^1_{\sqrt{2}} \, / \,  {\scriptstyle \left\lb \cT_{2\pi \frac{1}{\sqrt{2}}} \otimes
		  (-1)^{F^S_L} \right\rb},
\label{orbifold SU(2)_2}
\end{eqnarray}
where $F^S_L$ is the space-time fermion number associated with the left mover. 
Modding out by $\cT_{2\pi \frac{1}{\sqrt{2}}} \otimes (-1)^{F^S_L}$ makes
the NS-NS (R-R) sector to have even (odd) KK momenta. 
%%%
After incorporating suitable twisted sectors, 
%%%
this aligns the spin structures of the three fermions.
We then obtain the diagonal modular invariant of $SU(2)_2$ WZW\footnote
%%%
{The orbifolding \eqn{orbifold SU(2)_2} is quite similar 
to the Scherk-Schwarz compactification \cite{Scherk:1978ta} (or the
thermal superstring theory \cite{Atick:1988si}). In fact, the $SU(2)_2$
theory is useful in working with the thermal circle with inverse
temperature $\beta = 2\pi \sqrt{2} k $ ($k \in {\B Z}_{>0}$) 
in the RNS superstring, 
as discussed {\em e.g.} in \cite{Sugawara:2003xt}.
}
%%%
:
\beq
Z(\tau,\bar{\tau})=
\sum_{\ell=0,1,2}\left| \chi_\ell^{(2)}(\tau)\right|^2
=\half\Big(\left|\frac{\theta_2(\tau)}{\eta(\tau)}\right|^3
+\left|\frac{\theta_3(\tau)}{\eta(\tau)}\right|^3
+\left|\frac{\theta_4(\tau)}{\eta(\tau)}\right|^3\Big).
\label{Z SU(2)_2}
\eeq

As in the bosonic case, we define the T-fold as an orbifold generated by a group of order 2, 
namely the half-shift of the base combined with improved T-duality transformation:
\beq
T'':\;\;\;
X_L\rightarrow X_L+\pi\sqrt{\frac{1}{2}},\;\;\;
X_R\rightarrow -X_R,\;\;\;
\psi^X_L\rightarrow \psi^X_L,\;\;\;
\psi^X_R\rightarrow -\psi^X_R,
\label{eqn:superT2}
\eeq
which acts on the $SU(2)_2$ currents as
\beq
(J^1, J^2, J^3)\rightarrow (-J^1, -J^2, J^3)~, ~~~
(\tJ^1, \tJ^2, \tJ^3)\rightarrow (\tJ^1, -\tJ^2, -\tJ^3).
\eeq 
The transformation (\ref{eqn:superT2}) is again identified with asymmetric chiral rotation
\beq
T''=(e^{i\pi J^3_0}, e^{i\pi \bar J^1_0}).
\eeq
It is then straightforward to proceed as in the bosonic case to find the closed and open string spectra.
Instead of investigating this particular model, we shall in the next section explore a more general class of T-fold models with $SU(2)$ fibre at arbitrary level, which includes the $\cN=1$ T-fold as a special case at level 2.

%%%%%%%%%%%%%%%%%%%%%%%%%%%%%%%%%%%%%%%%%%%%%%%%%%%

Finally, we remark on application of the $\cN=1$ T-fold 
to models of superstring vacua.
For such purposes we need to generalise 
the fibre of the T-fold to a torus of even dimensions 
so that the chirality of the space-time fermions is unchanged under the T-duality action. 
A complication is that we need to carefully take account 
of the spin structures of the world-sheet fermions and the GSO condition, 
leading us to consider {\em truly} asymmetric modular invariants. 
This is certainly a very interesting subject and related work appeared in 
\cite{Brunner:1999fj,Gaberdiel:2002jr,Hellerman:2006tx}.
We hope to report on progresses in a separate publication.

%%%%%%%%%%%%%%%%%%%%%%%%%%%%%%%%%%%%%%%%%%%%%%%%%%%
%%%%%%%%%%%%%%%%%%%%%%%%%%%%%%%%%%%%%%%%%%%%%%%%%%%

%%%%%
\section{Extension to $SU(2)_k$ fibre}
%%%%%

In the above examples the $SU(2)$ structure was essential for obtaining the modular invariant one-loop partition functions and also for the existence of consistent boundary states.
As the bosonic and ${\C N}=1$ supersymmetric T-folds correspond to $SU(2)_k$ fibre with $k=1$ and $k=2$, it is natural to extend them to $SU(2)_k$ fibre of arbitrary level $k$. 
In this section we discuss such an extension.

%%%
\subsection{$SU(2)_k$ WZW T-fold}
%%%

The T-fold we shall consider  consists 
of the fibre of $SU(2)_k$ WZW model and the base which is a circle of radius $R$.
This is formulated as an orbifold
\beq
\left\lb SU(2)_k \times S^1_{2R} \right\rb/{\B Z}_2~,
\eeq
with the ${\B Z}_2$ orbifold action 
$\sigma \equiv (e^{i\pi J^3_0}, e^{i\pi \bar J^1_0})\otimes {\C T}_{2\pi R}$.
We shall be interested in the case where the fibre $SU(2)_k$ CFT is diagonal. 
As before, ${\C T}_{2\pi R}$ is the translation 
along the covering space of the base circle
${\C T}_{2\pi R}~:~Y\,\rightarrow \, Y+2\pi R$, 
and the $SU(2)_k$ currents are 
$J^a$ and $\bar J^a$, with $a=1,2,3$.
Twisting by $e^{i\pi J^3_0}$ or $e^{i\pi\bar J^1_0}$ 
generates a ${\B Z}_2$-orbifold of the chiral WZW model. 
The one-loop partition function of the T-fold is obtained from those of
the ${\B Z}_2$ WZW orbifolds and the base part, 
suitably combined in accordance with the T-invariant projection:   
\beq
Z^{\msc{$SU(2)$ T-fold}}(\tau,\bar{\tau}) 
= \frac{1}{2} \sum_{\alpha,\beta \in {\B Z}_2}\, 
Z^{\msc{base}}_{[\alpha,\beta]}(\tau,\bar{\tau})
\, \sum_{\ell=0}^k \chi^{(k)}_{\ell,[\alpha,\beta]}(\tau)
\overline{\chi^{(k)}_{\ell,[\alpha,\beta]}(\tau)}~.
\label{eqn:partSU(2)fiber}
\eeq
The definition and related formulas of the twisted $SU(2)$ characters
$\chi^{(k)}_{\ell, [\alpha,\beta]}(\tau)$ are summarized 
in Appendix \ref{sec:twistedSU(2)}. 
%%%
The modular invariant is again left-right symmetric, because the
$e^{i\pi J^3_0}$-twist (on the left-mover) and the $e^{i\pi \bar{J}^1_0}$-twist 
(the right-mover) result in the same character functions 
$\chi^{(k)}_{\ell, [\alpha,\beta]}$. 
%%%

To clarify the modular properties of the partition function
(\ref{eqn:partSU(2)fiber}) 
it is more convenient to use another notation 
of twisted characters $\widehat\chi^{(k)}_{\ell, (a,b)}(\tau)$, defined in 
(\ref{eqn:halftwistedSU(2)chars}).
These differ from $\chi^{(k)}_{\ell, [\alpha,\beta]}(\tau)$ 
only by phase normalisation and are
covariant under modular transformations
(see \eqn{eqn:T-twistedSU(2)char} and \eqn{eqn:S-twistedSU(2)char}).
One may then rewrite the partition function as
\begin{eqnarray}
 Z^{\msc{$SU(2)$ T-fold}}(\tau,\bar{\tau}) &=&
\frac{1}{2} \sum_{\alpha,\beta \in {\B Z}_2}\, 
Z^{\msc{base}}_{[\alpha,\beta]}(\tau,\bar{\tau})
\, \sum_{\ell=0}^k \widehat\chi^{(k)}_{\ell,(\alpha/2,\beta/2)}(\tau)
\overline{\widehat\chi^{(k)}_{\ell,(\alpha/2,\beta/2)}(\tau)} \nn\\
&=& 
\sum_{w,m\in {\B Z}}\, 
Z_{R, (w,m)}(\tau,\bar{\tau})
\, \sum_{\ell=0}^k \widehat\chi^{(k)}_{\ell,(w/2,m/2)}(\tau)
\overline{\widehat\chi^{(k)}_{\ell,(w/2,m/2)}(\tau)}~,
\label{eqn:partSU(2)fiber2}
\end{eqnarray}
where $Z_{R,(w,m)}(\tau,\bar\tau)$ is defined in (\ref{eqn:bosonZ}).
This is manifestly modular invariant, since each piece behaves covariantly
under modular transformations.

%%%%%%%%%%%%%%%%%%%%%%%%%%%%%%%%%%%%%%%%%%%%%%%%%%%

We incidentally remark that if merely the modular invariance is concerned,  
another (entirely asymmetric) modular invariant is possible: 
\beq
 Z'(\tau,\bar{\tau}) =\sum_{w,m\in {\B Z}}\, 
Z_{R, (w,m)}(\tau,\bar{\tau})\, \sum_{\ell=0}^k \chi^{(k)}_{\ell}(\tau)
\overline{\widehat\chi^{(k)}_{\ell,(w/2,m/2)}(\tau)}~.
\label{eqn:SU(2)asymmetricZ}
\eeq
Since this is generated by an asymmetric action ${\C T}_{2\pi R} \otimes ({\bf 1}, e^{i\pi \bar J^1_0})$ 
that does not contain $e^{i\pi J^3_0}$-twist on the fibre, 
it may be regarded as a T-fold with the original definition of T-duality 
($T$, without the $X_L$-translation nor the phase shift $e^{i\pi\hat n\hat w}$).
While modular invariant by construction, whether this model has any relevance as a physically acceptable string vacuum is not immediately clear to us. 
There is level mismatch in the twisted sectors in general, and the model is not an orbifold of order 2.
The order of the orbifold group is
$N\equiv \mbox{L.C.M} \, \{N',2\}$, 
where $N'$ is the smallest positive integer such that 
$e^{2\pi i \frac{N' k}{16}}=1$. 
In the level $k=1$ case (a bosonic T-fold of $S^1$-fiber), for instance, this construction gives rise to
an asymmetric modular invariant of an order 16 orbifold.
Similarly, $k=2$ (an $\cN=1$ T-fold of $S^1$-fiber) leads to an order 8 asymmetric orbifold. 
In those cases, unfortunately, there arises a problem of locality of vertex operators. 
Below in this section we shall focus on the model given by (\ref{eqn:partSU(2)fiber}).

%%%%%%%%%%%%%%%%%%%%%%%%%%%%%%%%%%%%%%%%%%%%%%%%%%%
%%%%%%%%%%%%%%%%%%%%%%%%%%%%%%%%%%%%%%%%%%%%%%%%%%%

%%%
\subsection{Bulk branes in the $SU(2)$ T-fold}
%%%

Let us consider $SU(2)_k$ generalisation of the bulk branes discussed in
Sec.\ref{sec:bulkB}. We shall first focus on the familiar Cardy states 
\cite{Cardy:1989ir} defined by  ($L=0,1,\ldots,k$)
\begin{eqnarray}
 && \ket{L}_C \equiv \sum_{\ell =0}^k \, \frac{S^{(k)}_{L,\ell}}
{\sqrt{S^{(k)}_{0,\ell}}} \, \dket{\ell}, 
%\nn \\
%&& (J^a_n+\tJ^a_{-n})\ket{L}_C=0~, ~~~ ({}^{\forall}n, ~ {}^{\forall}
%a)~,
\label{Cardy SU(2)}
\end{eqnarray}
where $S^{(k)}_{\ell,\ell'} \equiv \sqrt{\frac{2}{k+2}}
 \sin \left(\pi\frac{(\ell+1)(\ell'+1)}{k+2}\right)$ is the modular S-matrix
 of $SU(2)_k$, and the Ishibashi states \cite{Ishibashi:1988kg} $\dket{\ell}$
 are characterized by
 \begin{eqnarray}
 && (J^a_n+\bar J^a_{-n}) \dket{\ell} = 0, ~~~ ({}^{\forall}n, ~
  {}^{\forall}a),  
\label{Ishibashi SU(2)}\\
 && \dbra{\ell} e^{-\pi s H^{c}} e^{2\pi i z J^3_0}\dket{\ell'} 
= \delta_{\ell,\ell'} \chi^{(k)}_{\ell}(z|is).
\label{overlap Ishibashi SU(2)}
\end{eqnarray}
In this expression $\chi^{(k)}_{\ell}(z|is)$ is the $SU(2)_k$ character of spin
$\ell/2$ \eqn{eqn:SU(2)char},
and $H^{c}\equiv L_0+\tL_0 -\frac{c_k}{12}$ is the closed string
Hamiltonian.
It is well-known that these `maximally symmetric' boundary states
$\ket{L}_C$ describe D-branes wrapped on the conjugacy classes of 
$SU(2)$,
interpreted as $(k-1)$ spherical D2 branes (for $L=1,\ldots,k-1$)\footnote 
%%%
{
A D$p$-brane in our context is a $p$-dimensional object spreading in $p$ spatial dimensions 
(not in $(p+1)$ spacetime dimensions).
}
%%%
and two D0 particles at the poles of $S^3$ ($L=0,k$)
\cite{Alekseev:1998mc}.

We also introduce `T-dualized' boundary states associated to 
$T'' \equiv (e^{i\pi J^3_0}, e^{i\pi \tJ^1_0})$,
\begin{eqnarray}
 && \hket{L}_C \equiv T'' \ket{L}_C =
\sum_{\ell =0}^k \frac{S^{(k)}_{L,\ell}}
{\sqrt{S^{(k)}_{0,\ell}}} \, \hdket{\ell}, 
\label{Cardy SU(2) T} \\
&& \hdket{\ell} \equiv T'' \dket{\ell} = e^{i\pi
 J^2_0}\dket{\ell}.
\label{Ishibashi SU(2) T}
\end{eqnarray}
which satisfy
\begin{eqnarray}
 && (J^3_n-\tJ^3_{-n})\hket{L}_C=0, 
~~~ (J^{\pm}_n-\tJ^{\mp}_{-n})\hket{L}_C=0
\end{eqnarray}
%and has the overlaps 
%\begin{eqnarray}
% \hdbra{\ell} e^{-\pi s H^{(c)}} e^{2\pi i z J^3_0}\dhket{\ell'} 
%&=& \delta_{\ell,\ell'} \chi^{(k)}_{\ell}(z|is), \nn\\
%  \hdbra{\ell} e^{-\pi s H^{(c)}} e^{2\pi i z J^3_0}\dket{\ell'}
%&=&  \dbra{\ell} e^{-\pi s H^{(c)}} e^{2\pi i z J^3_0}\hdket{\ell'}
%= \delta_{\ell,\ell'} \hchi^{(k)}_{\ell}(z|is)
%\end{eqnarray}
(note that $e^{i\pi J^3_0}e^{-i\pi J^1_0} = e^{i\pi J^2_0}$).

Using the overlaps \eqn{overlap Ishibashi SU(2)} and 
the Verlinde formula
\begin{eqnarray}
 \frac{S^{(k)}_{L_1,\ell}S^{(k)}_{L_2,\ell} }{S^{(k)}_{0,\ell}}
= \sum_{L=0}^k\, N_{L_1,L_2}^{L} S^{(k)}_{L,\ell}~,
\end{eqnarray}
where $N^{L}_{L_1,L_2}$ denotes the fusion coefficients of $SU(2)_k$,
it is easy to evaluate the cylinder amplitudes as 
\begin{eqnarray}
\hspace{-1cm}
 {}_C\bra{L_1} e^{-\pi s H^{c}} \ket{L_2}_C &=& 
  {}_C\hbra{L_1} e^{-\pi s H^{c}} \hket{L_2}_C
% \nn \\
= \sum_{L=0}^k \,  N_{L_1,L_2}^{L} \chi^{(k)}_L(0|it)
\equiv Z_{SU(2)_k}^{L_1,L_2}(it)
. \nn \\
%%%
\hspace{-1cm}
 {}_C\bra{L_1} e^{-\pi s H^{c}} \hket{L_2}_C &=& 
  {}_C\hbra{L_1} e^{-\pi s H^{c}} \ket{L_2}_C 
%\nn \\
= \sum_{L=0}^k \,  N_{L_1,L_2}^{L} \chi^{(k)}_{L,[1,0]}(0|it)
\equiv \hZ_{SU(2)_k}^{L_1,L_2}(it)~.
\label{cylinder SU(2) 1} 
\end{eqnarray}
Here $t\equiv 1/s$ is the open string modulus of the cylinder. 
%$\chi^{(k)}_L(0|it)$,
$ \chi^k_{L,[1,0]}(it)$ 
are the twisted $SU(2)_k$ characters 
given in \eqn{eqn:twistedSU(2)char}).

Now, the bulk branes are constructed  similarly to \eqn{eqn:bulkDD}, \eqn{eqn:bulkNN},
\begin{eqnarray}
 \ket{L,D(y_0)} &=& \frac{1}{\sqrt{2}}(1+\sigma) \ket{L}_C
  \otimes \ket{D(y_0)}_{2R} \nn\\
&\equiv & \frac{1}{\sqrt{2}} \left(
\ket{L}_C \otimes \ket{D(y_0)}_{2R} + 
\hket{L}_C \otimes \ket{D(y_0+2\pi R)}_{2R}
\right), 
\label{eqn:bulkSU(2)D} \\
%%%
 \ket{L,N(\ty_0)} &=& \frac{1}{\sqrt{2}}(1+\sigma) \ket{L}_C
  \otimes \ket{N(\ty_0)}_{2R} \nn\\
&\equiv & \frac{1}{\sqrt{2}}
\left(\ket{L}_C + \hket{L}_C\right) \otimes \ket{N(\ty_0)}_{2R},
\label{eqn:bulkSU(2)N}
\end{eqnarray}
where $\sigma \equiv T^{''}\otimes\cT_{2\pi R}$.
The overlaps between the bulk branes are computed as 
\begin{eqnarray}
 && \hspace{-1cm}
\langle L, D(y_0)\vert e^{-\pi sH^c}\vert L', D(y'_0)\rangle
%\nn\\&&
%\hspace{35mm}
=Z_{SU(2)_k}^{L,L'}(it)Z^{DD}_{2R}(it;\Delta y_0)
+\hZ_{SU(2)_k}^{L,L'}(it)Z^{DD}_{2R}(it;\Delta y_0+2\pi R),\nn\\
%%%
&& \hspace{-1cm}
\langle L, N(\tilde y_0)\vert e^{-\pi sH^c}\vert L', N(\tilde y'_0)
\rangle
%\nn\\&&
=\Big(Z_{SU(2)_k}^{L,L'}(it)+\hZ_{SU(2)_k}^{L,L'}(it)\Big) 
Z^{NN}_{2R}(it;\Delta\tilde y_0),\nn\\
%%%
&& \hspace{-1cm}
\langle L, D(y_0)\vert e^{-\pi sH^c}\vert L', N(\tilde y'_0)\rangle
%\nn\\&&
=\Big(Z_{SU(2)_k}^{L,L'}(it)+\hZ_{SU(2)_k}^{L,L'}(it)\Big) Z^{DN}(it).
\label{cyl amp SU(2) T-fold}
\end{eqnarray}
Again these have obvious geometrical interpretation on the
universal cover of the base $S^1$.

This construction may be generalised to include marginal boundary deformation 
by an arbitrary $SU(2)$-rotation on the fibre. 
Such deformation is taken into account by replacing the Cardy states $\ket{L}_C$ 
along the $SU(2)$-fiber with the deformed Cardy states,
\begin{eqnarray}
 && \ket{L,\om}_C \equiv \bar{R}(\om) \ket{L}_C \left(\equiv 
R(\om^{-1}) \ket{L}_C\right),
\label{rotated Cardy SU(2)}
\end{eqnarray}
where the rotations are defined by
%\begin{eqnarray}
% && R(\om) \equiv \prod_{j}e^{i\theta_{j}J^{a_j}_0}, ~~~
%\bar{R}(\om)\equiv \prod_{j}e^{i\theta_{j}\tJ^{a_j}_0},
%\end{eqnarray}
\beq
R(\om) \equiv \exp\sum_a{i\theta_{a}J^{a}_0}, ~~~
\bar{R}(\om)\equiv \exp\sum_a{i\theta_{a}\tJ^{a}_0},
\eeq
with 
${}^{\forall}\om \equiv\exp\sum_a{i\theta_{a}\frac{\sigma_{a}}{2}}
\in SU(2)$
%${}^{\forall}\om \equiv \prod_j e^{i\theta_{j}\frac{\sigma_{a_j}}{2}}
%\in SU(2)$
($\sigma_a$ are the Pauli matrices).
This type of boundary states is characterized by twisted gluing conditions:
\begin{eqnarray}
 (J^a_n+ \Ad(\om)_{ba}\tJ^b_{-n}) \ket{L,\om}_C=0. ~~~ ({}^{\forall}a,~~ 
{}^{\forall}n).
\label{twisted gluing cond}
\end{eqnarray}
Then the bulk branes are,
\begin{eqnarray}
  \ket{(L,\om),D(y_0)}^{\rm bulk} &=& \frac{1}{\sqrt{2}}(1+\sigma) \ket{L,\om}_C
  \otimes \ket{D(y_0)}_{2R},~~~ \mbox{etc.}
\end{eqnarray}
and the overlaps are calculable by means of the diagonalization technique described 
in Appendix \ref{sec:MixedFormulae}. 
We find, for instance,
\begin{eqnarray}
&&\hspace{-10mm}\langle (L,\om), D(y_0)\vert e^{-\pi sH^c}\vert (L',\om'), D(y'_0)\rangle\nn\\
&&\hspace{-10mm}
=Z_{SU(2)_k}^{L,L'}(it;\xi(\om\om^{' -1}))Z^{DD}_{2R}(it;\Delta y_0)
+Z_{SU(2)_k}^{L,L'}(it; \xi\left(\om e^{\frac{i\pi}{2}\sigma_3}\om^{' -1}
e^{-\frac{i\pi}{2}\sigma_1}\right)
)Z^{DD}_{2R}(it;\Delta y_0+2\pi R), \nn\\
\label{cyl SU(2) T-fold twisted}
\end{eqnarray}
where
\begin{eqnarray}
 && Z_{SU(2)_k}^{L_1,L_2}(it;z) \equiv \sum_L N_{L_1,L_2}^L \chi_L^{(k)}
\left(it z|it \right) e^{-\frac{\pi k}{2}t z^2} , 
\end{eqnarray}
and $\xi(\om) \in [0,1]$ 
($\om \in SU(2)$) is defined by diagonalization 
\begin{eqnarray}
 U e^{2\pi i\xi(\om)\frac{\sigma_3}{2}} U^{-1} 
= \om, ~~~ \mbox{with some }U \in SU(2).
\end{eqnarray}
In the particular case of
$\om= e^{i\theta \sigma_3}$, $\om'= e^{i\theta' \sigma_3}$
we obtain 
\begin{eqnarray}
&& 
\hspace{-5mm}
\langle (L,\theta), D(y_0)\vert e^{-\pi sH^c}\vert (L',\theta'), 
D(y'_0)\rangle
\nn\\&&
\hspace{1cm}
=Z_{SU(2)_k}^{L,L'}(it;\frac{\theta-\theta'}{\pi})Z^{DD}_{2R}(it;\Delta y_0)
+\hZ_{SU(2)_k}^{L,L'}(it)Z^{DD}_{2R}(it;\Delta y_0+2\pi R).
\label{cyl SU(2) T-fold twisted 2}
\end{eqnarray}
Other overlaps are evaluated in the same way.

%%%%%%%%%%%%%%%%%%%%%%%%%%%%%%%%%%%%%%%%%%%%%%%%%%%

%
\subsection{Fractional branes in the $SU(2)$ T-fold}

The fractional branes \eqn{eqn:FracBrane1} may also be generalised to 
$SU(2)_k$ fibre at arbitrary level.
Their boundary states are found to be
%%%
\begin{eqnarray}
 && \hspace{-1cm}
\ket{F;(L,\theta),\ty_0, \eta=\pm 1 } \equiv \frac{1}{\sqrt{2}}e^{2i\theta
  J^3_0}\kappa \ket{L}_C \otimes \ket{N(\ty_0)}_{2R}
+ \eta \frac{1}{\sqrt{2}}e^{2i\theta
  J^3_0} \kappa \ket{L}^{\cR}_C \otimes \ket{N(\ty_0)}^{\cT}_{2R}.
\label{frac brane SU(2) T-fold}
\end{eqnarray}
Here, $\kappa$ is the same automorphism \eqn{kappa} as before but now
for $SU(2)_k$. 
$\ket{L}_C$ are the $SU(2)_k$ Cardy states \eqn{Cardy SU(2)} and $\ket{L}_C^{\cR}$ 
are their twisted counterparts, defined explicitly as 
\begin{eqnarray}
&& \ket{L}_C^{\cR} \equiv \sum_{\ell=0}^k 
\frac{e^{\frac{i\pi }{2}L}S^{(k)}_{L,\ell}}{\sqrt{S^{(k)}_{0,\ell}}} 
\dket{\ell}^{\cR}, 
\label{Cardy SU(2) twisted} \\
%%%
&& (J^1_n+\tJ^1_{-n})\dket{\ell}^{\cR}=0, ~~~ ({}^{\forall}n \in {\B Z}),
  \nn\\
&& (J^a_r+\tJ^a_{-r})\dket{\ell}^{\cR}=0, ~~~ ({}^{\forall}r \in
  \frac{1}{2}+{\B Z},~~
a=2,3), \nn\\
&& {}^{\cR}\dbra{\ell}e^{-\pi s H^{(c)}} e^{2\pi i z J^1_0}
 \dket{\ell'}^{\cR}
= 
%\delta_{\ell,\ell'} \hchi_{\ell, (1/2,0)}(z|is) 
\delta_{\ell,\ell'} \chi^{(k)}_{\ell, [1,0]}(z|is).
\label{Ishibashi SU(2) twisted}
\end{eqnarray}
The necessity of the slightly non-trivial phase factor $e^{i\frac{\pi}{2}L}$
will be clarified below.
The states in the base part $\ket{N(\ty)}_{2R}$, $\ket{N(\ty)}_{2R}^{\cT}$ 
are exactly same as before. 
The construction and analysis of the fractional states heavily rely on various properties of
the ${\B Z}_2$-twisted $SU(2)_k$ characters $\chi^{(k)}_{\ell,[\al,\beta]}(z|\tau)$ ($\al,\beta \in {\B Z}_2$).
See Appendix \ref{sec:twistedSU(2)} for their definitions and properties.
The periodicity of the continuous marginal deformation parameter $\theta$ 
is summarized as follows:
\begin{description}
 \item [(i) $k$ : even ]

~

The periodicity of $\theta$ is $2 \pi $:
\begin{eqnarray}
 && \ket{F; (L,\th+2\pi), \ty_0, \eta} = \ket{F; (L,\th), \ty_0, \eta},
\label{periodicity k even}
\end{eqnarray}
and we must treat $\ket{F;(L,\th),\ty_0,+}$ and 
$\ket{F;(L,\th),\ty_0,-}$ independently. 
We also note 
\begin{eqnarray}
 && \ket{F; (L,\th+\pi), \ty_0, \eta} = \ket{F; (k-L,\th), \ty_0, (-1)^L\eta}.
\label{rel L k-L}
\end{eqnarray}
%%%
 \item [(ii) $k$ : odd ]

~

The periodicity of $\theta$ is $4 \pi $:
\begin{eqnarray}
 && \ket{F; (L,\th+4\pi), \ty_0, \eta} = \ket{F; (L,\th), \ty_0, \eta},
\label{periodicity k odd}
\end{eqnarray}
and $\ket{F;(L,\th),\ty_0,+}$ and $\ket{F;(L,\th),\ty_0,-}$ are related as 
\begin{eqnarray}
 && \ket{F; (L,\th+2\pi), \ty_0, \eta} = \ket{F; (L,\th), \ty_0, -\eta}.
\end{eqnarray}
We again obtain the same relation as \eqn{rel L k-L} when shifting 
$\theta\,\rightarrow\, \theta+\pi$.
%We also note 
%\begin{eqnarray}
% && \ket{F; (L,\th+\pi), \ty, \eta} = \ket{F; (k-L,\th), \ty, (-1)^L\eta}.
%\end{eqnarray}
%
\end{description}

%%%%%%%%%%%%%%%%%%%%%%%%%%%%%%%%%%%%%%%%%%%%%%%%%%%

Computation of the cylinder amplitudes is carried out in the same way as in
the $S^1$-fiber T-fold. 
With the help of modular transformation formulas 
of the twisted characters \eqn{eqn:modulartwistedSU(2)-2}, we find 
\begin{eqnarray}
&& \hspace{-30mm} \bra{F;(L_1,\th),\ty_0, \eta} 
e^{-\pi s H^{(c)}} \ket{F;(L_2,\th'),\ty'_0,\eta'} \nn \\
&& \hspace{-10mm}
= \frac{1}{\eta(it)} \sum_{n\in 2{\B Z}}e^{-2\pi t
\left(\frac{n}{2R}+\frac{\Delta \ty_0}{2\pi}\right)^2}
\, \sum_{L}N^L_{L_1,L_2}\chi^{(k)}_L
(it \frac{\Delta \th}{\pi}| it ) e^{-\frac{kt}{2\pi} (\Delta \th)^2} 
\nn \\
&& \hspace{-10mm}
+ \frac{\eta\eta'}{\eta(it)} \sum_{n\in 2{\B Z}+1 }e^{-2\pi t
\left(\frac{n}{2R}+\frac{\Delta \ty_0}{2\pi}\right)^2}
\, \sum_{L}N^L_{L_1,L_2}e^{-i\frac{\pi}{2}(L_1-L_2+L)}
\chi^{(k)}_{L,[0,1]}(it \frac{\Delta \th}{\pi}|
it ) e^{-\frac{kt}{2\pi} (\Delta \th)^2},
\label{cyl amp frac 1}
\\
%%%
&& \hspace{-30mm}\bra{(L_1,\th), N(\ty_0)} e^{-\pi s H^{(c)}}
\ket{F;(L_2,\th'),\ty'_0,\eta'}
= Z_{2R}^{NN}(it; \Delta \ty_0) \sum_L N_{L_1,L_2}^L 
\chi^{(k)}_L (it \frac{\al(\Delta \th)}{\pi}| it)
e^{-\frac{kt}{2\pi} \al(\Delta \th)^2}, 
\label{cyl amp frac 2}
\\
%%%
&& \hspace{-30mm}\bra{(L_1,\th), D(y_0)} e^{-\pi s H^{(c)}}
\ket{F;(L_2,\th'),\ty'_0,\eta'}
= Z^{DN}(it) \sum_L N_{L_1,L_2}^L 
\chi^{(k)}_L (it \frac{\al(\Delta \th)}{\pi}| it)
e^{-\frac{kt}{2\pi} \al(\Delta \th)^2},
\label{cyl amp frac 3}
\end{eqnarray}
where
$\al(\th)\equiv \cos^{-1}\left(\frac{\cos \th}{\sqrt{2}}\right)$, 
$\Delta \th \equiv \th-\th'$ and $\Delta \ty_0 \equiv \ty_0 -\ty'_0$
as before.
%%%
% Comments on the branes in the $SU(2)_k$ T-fold
%%%
We would like to conclude this section with several comments on these branes.

\noindent
{\bf 1. } 
A non-trivial point is the inclusion of the phase factor 
$e^{i\frac{\pi}{2}L}$ in 
\eqn{Cardy SU(2) twisted}. 
This factor is indeed necessary for an appropriate ${\B Z}_2$-projection 
in the open string channel. 
Without this factor, the open channel amplitude would be twisted by 
$e^{i\pi J^2_0}$ which is not involutive: 
$(e^{i\pi J^2_0})^2= e^{2\pi i J^2_0} \neq {\bf 1}$.
%\footnote
%{
%	On the other hand, the operator $(e^{i\pi J^3_0}, e^{i\pi \tJ^1_0})$
%	is involutive on the closed string Hilbert space, because 
%	unwanted phase factors are cancelled out between the left and right movers.
%  }. 
See also Appendix \ref{sec:twistedSU(2)}.
We also note that $e^{i\frac{\pi}{2}(L_1-L_2+L)} = \pm 1$, because 
$L_1-L_2+L \in 2{\B Z}$ when $N_{L_1,L_2}^L \neq 0$. 
Therefore, \eqn{cyl amp frac 1} is correctly ${\B Z}_2$-projected and the Cardy
condition is satisfied among the boundary states we defined.

%%%%%%%%%%%%%%%%%%%%%%%%%%%%%%%%%%%%%%%%%%%%%%%%%%%

\noindent
{\bf 2. } An alternative way to construct the boundary states of the
fractional branes is to focus on the primary states of the orbifold 
$SU(2)_k/{\B Z}_2$. 
To this aim it is helpful to recall the level 1 case which was elaborated 
in Sec. \ref{sec:fracB}. 
We have 8 primary states corresponding to irreducible characters
\bea
&&\chi_{\B I}(\tau)=\half \big(\chi^{(1)}_0(\tau)+\chi^{(1)}_{0,[0,1]}(\tau)\big),\nn\\
&&\chi_{j}(\tau)=\half \big(\chi^{(1)}_0(\tau)-\chi^{(1)}_{0,[0,1]}(\tau)\big),\nn\\
&&\chi_{1}^{i}(\tau)=\half \big(\chi^{(1)}_1(\tau)\pm
\chi^{(1)}_{1,[0,1]}(\tau)\big)
\left(\, \equiv \half \chi^{(1)}_1(\tau) \, \right)
,\nn\\
&&\chi_{\sigma}^{i}(\tau)=\half \big(\chi^{(1)}_{\ell,[1,0]}(\tau)+\chi^{(1)}_{\ell,[1,1]}(\tau)\big),\nn\\
&&\chi_{\tau}^{i}(\tau)=\half \big(\chi^{(1)}_{\ell,[1,0]}(\tau)-\chi^{(1)}_{\ell,[1,1]}(\tau)\big),
\eea
where $i=1,2$, $\ell=0,1$ and the notations are as in
Sec.\ref{sec:fracB}
(see also Appendix \ref{sec:RCFT} and \ref{sec:twistedSU(2)}).
These characters are organized into a collective form,
\beq
\chi_{\ell}^{[\alpha], \pm}(\tau)
=\half \big(\chi^{(1)}_{\ell,[\alpha,0]}(\tau)\pm\chi^{(1)}_{\ell,[\alpha,1]}(\tau)\big).
\eeq
Here, $\alpha=0,1$ and $\chi^{(1)}_{\ell,[0,0]}(\tau)\equiv\chi^{(1)}_{\ell}(\tau)$.
We write them as $\chi_{\ell}^{s}(\tau)$, with $s=([0],+)$, 
$([0],-)$, $([1],+)$, $([1],-)$ in this order.
The correspondence to the ${\C A}_1/{\B Z}_2$ labels is as shown in Table 2.
%%%%%
%Table 2
%%%%%
\begin{center}
\begin{tabular}{c||ccc|cc}
%\hline
% after \\ : \hline or \cline{col1-col2} \cline{col3-col4} ...
Sector &\multicolumn{3}{c|}{Untwisted}&\multicolumn{2}{c}{Twisted}\\
\hline
Conformal weight & $0$ & $1$ & $\quarter$ & $\frac{1}{16}$ &$\frac{9}{16}$\\
\hline
${\C A}_1/{\B Z}_2$ primary 
& ${\B I}$ & $j$ & $\phi^i_1$ & $\sigma^i$ & $\tau^i$\\
$\ell$ & $0$ & $0$ & $1$ & $0,1$ & $0,1$ \\
$s$ & $[0],+$ & $[0],-$ & $[0],\pm$ & $[1],+$ & $[1],-$\\
%\hline
\end{tabular}\\
\vspace{10pt}
{\bf Table 2.} Primary fields in $SU(2)_1/{\B Z}_2$.
\end{center}
%%%
%
%
This rational CFT is generalised to $SU(2)_k/{\B Z}_2$ with arbitrary $k$\cite{Kac:1996nq,Birke:1999ik}.
A natural generalisation of the character formulas is
\beq
\chi_{\ell}^{[\alpha], \pm}(\tau)=\half \big(\chi^{(k)}_{\ell,[\alpha,0]}(\tau)\pm\chi^{(k)}_{\ell,[\alpha,1]}(\tau)\big),
\label{ch SU(2)_k/Z_2}
\eeq
with now $\ell=0, 1, \cdots, k$.
Note that the diagonal sum
$
\sum_{\ell,s}\left|\chi_\ell^s(\tau)\right|^2
$
of the $4(k+1)$ characters \eqn{ch SU(2)_k/Z_2} gives the fibre part of the partition function
\eqn{eqn:partSU(2)fiber}.
Modular inversion of these characters are  
%(we denote $([0],+)$, $([0],-)$, $([1],+)$, $([1],-)$ 
%as $s=1,2,3,4$ in this order);
\beq
\chi_{\ell}^{s}(-1/\tau)=
\sum_{\ell',s'}\, 
%\cS_{(\ell,s), (\ell',s')}\,
S^{(k)}_{\ell,\ell'} M^{(\ell,\ell')}_{s,s'}\,
\chi_{\ell'}^{s'}(\tau),
\eeq
with
\begin{eqnarray}
% \cS_{(\ell,s), (\ell',s')} &=& S^{(k)}_{\ell,\ell'}
%  M_{s,s'}^{(\ell,\ell')}  , \\
M^{(\ell,\ell')}_{s,s'}&=&\half
\left[\begin{array}{cccc}1 & 1 & e^{i\frac{\pi}{2}\ell}
 &e^{i\frac{\pi}{2}\ell}  \\
 1 & 1 & -e^{i\frac{\pi}{2}\ell} & -e^{i\frac{\pi}{2}\ell} \\
 e^{i\frac{\pi}{2}\ell'} & -e^{i\frac{\pi}{2}\ell'} & e^{\frac{\pi
  i}{2}(\ell+\ell'-\frac k2)} & -e^{\frac{\pi i}{2}(\ell+\ell'-\frac
  k2)} \\
e^{i\frac{\pi}{2}\ell'} & -e^{i\frac{\pi}{2}\ell'} & 
-e^{\frac{\pi i}{2}(\ell+\ell'-\frac k2)} & e^{\frac{\pi i}{2}(\ell+\ell'-\frac k2)}\end{array}\right].
\label{eqn:SU(2)k/Z2_modularS}
\end{eqnarray}
It is easy to check the unitarity of the modular matrix.
We can now construct the $4(k+1)$ Cardy states based on the
modular data \eqn{eqn:SU(2)k/Z2_modularS} following the standard procedure 
of boundary RCFT, yielding the fractional boundary states as in Sec. \ref{sec:fracB} 
(with the help of the automorphism $\kappa$). 
It is not difficult to see the $4(k+1)$ Cardy states found this way coincide (up to phase factors)
with $\ket{F;(L,\th),\ty_0,\pm}$ with values of the parameter $\th$ suitably chosen; 
we find correspondence
\begin{eqnarray}
 && L=0,1,\ldots, \left[\frac{k}{2}\right], ~~~ \th= \frac{n \pi}{2}, ~~ 
(n=0,1, \ldots, 3), ~~~ \eta=\pm 1, ~~~ \mbox{(for even $k$)}, \nn \\
 && L=0,1,\ldots, \left[\frac{k}{2}\right], ~~~ \th= \frac{n \pi}{2}, ~~ 
(n=0,1, \ldots, 7), ~~~ \eta=+ 1, ~~~ \mbox{(for odd $k$)}. 
\end{eqnarray} 
(Only half of the $L$-values are independent. Recall \eqn{rel L k-L}.) 
The factor $e^{i\frac{\pi}{2}L}$ in \eqn{Cardy SU(2) twisted} is again essential in this correspondence.
One can also easily check that the results in Sec. \ref{sec:fracB} are reproduced in the case of $k=1$.

~

%%%%%%%%%%%%%%%%%%%%%%%%%%%%%%%%%%%%%%%%%%%%%%%%%%%

\noindent
{\bf 3. } In the $SU(2)$ WZW there are also B-branes \cite{Maldacena:2001ky} that 
preserve only a part of the $SU(2)$ symmetry on the boundary and are
interpreted geometrically as D3-branes or (blown-up) D1-branes, not
corresponding to any conjugacy classes.
In our $SU(2)$ T-fold model it seems possible to construct bulk boundary states out of 
such B-type $SU(2)$ boundary states, although we have not developed 
them in full detail.
Exploration of such branes and investigation of completeness of D-branes 
(in the sense of \cite{Pradisi:1996yd}) are certainly intriguing problems and we hope to come 
back in our future work.  

~

%%%%%%%%%%%%%%%%%%%%%%%%%%%%%%%%%%%%%%%%%%%%%%%%%%%

\noindent
{\bf 4. } Finally, we would like to mention the model described by the asymmetric 
modular invariant \eqn{eqn:SU(2)asymmetricZ}.
As already pointed out this orbifold is somewhat pathological and it may not serve as
a sensible model of string background. 
Nevertheless the model is legitimate as a field theory and it is an interesting problem to look into the spectrum of D-branes.
The construction of bulk branes is essentially same as those discussed above; 
the corresponding boundary states are 
obtained by adding images of the orbifold action (which is not involutive in this case).
In contrast, fractional branes are absent in this orbifold since the conformal invariance on the boundary is broken in the twisted sectors (due to the level mismatch).
In similar but less simple examples of asymmetric orbifolds 
%%%
(associated with tori of higher dimensions), 
%%%
fractional-type branes are often possible 
due to cancellation of the level-mismatch, as observed in \cite{Brunner:1999fj,Gaberdiel:2002jr}.

%%%
%\section{Discussion}
%%%

\section*{Acknowledgments}

Y.S. was partly supported by Ministry of Education, Culture, Sports, Science and Technology of Japan.
S.K. acknowledges support from JSPS (Research Fellowship for Young Scientists) and the Academy of Finland (Finnish-Japanese Core Programme, grant 112420). 
It is our pleasure to thank for the hospitality of APCTP, Pohang, 
where this work was initiated
(2006 Focus Program: `Liouville, Integrability and Branes (3)').

%%%%%%%%%%%%%%%%%%%%%%%%%%%%%%%%%%%%%%%%%%%%%%%%%%%
%%%%%%%%%%%%%%%%%%%%%%%%%%%%%%%%%%%%%%%%%%%%%%%%%%%

\appendix

\section{Notations and Conventions}
\label{sec:convention}

We first summarize our convention of theta functions. 
We let $q\equiv e^{2\pi i \tau}$, $y\equiv e^{2\pi iz}$ and define
\beq
\Theta_{m,n}(z|\tau)=\sum_{k\in{\B Z}}
q^{n(k+\frac{m}{2n})^2}y^{n(k+\frac{m}{2n})},\;\;\;
\tilde\Theta_{m,n}(z|\tau)
=\sum_{k\in{\B Z }}(-1)^kq^{n(k+\frac{m}{2n})^2}y^{n(k+\frac{m}{2n})},
\label{Theta fn}
\eeq
%We also use the standard eta and theta functions
\begin{eqnarray}
 \eta(\tau)&=&q^{1/24}\prod_{n=1}^\infty(1-q^n),
\end{eqnarray}
\bea
%\eta(\tau)&=&q^{1/24}\prod_{n=1}^\infty(1-q^n),\nn\\
\theta_1(z\vert \tau)&=& 
-i\sum_{n\in{\B Z}}(-1)^{n}y^{n+\half}q^{\half(n+\half)^2}
\equiv 2 \sin(\pi z)q^{\eighth}\prod_{m=1}^{\infty}
    (1-q^m)(1-yq^m)(1-y^{-1}q^m)
,\nn\\
\theta_2(z\vert \tau)&=&\sum_{n\in{\B Z}}y^{n+\half}q^{\half(n+\half)^2}
\equiv  2 \cos(\pi z)q^{\eighth}\prod_{m=1}^{\infty}
    (1-q^m)(1+yq^m)(1+y^{-1}q^m)
,\nn\\
\theta_3(z\vert \tau)&=&\sum_{n\in{\B Z}}y^nq^{\half n^2}
\equiv \prod_{m=1}^{\infty}
    (1-q^m)(1+yq^{m-\half})(1+y^{-1}q^{m-\half})
,\nn\\
\theta_4(z\vert \tau)&=&\sum_{n\in{\B Z}}(-1)^ny^nq^{\half n^2}
    \equiv \prod_{m=1}^{\infty}
    (1-q^m)(1-yq^{m-\half})(1-y^{-1}q^{m-\half}),
\label{theta fn}
\eea
and we abbreviate as $\Th{m}{n}(\tau)\equiv \Th{m}{n}(0|\tau)$,
$\tTh{m}{n}(\tau)\equiv \tTh{m}{n}(0|\tau)$,
$\theta_i(\tau)=\theta_i(0\vert\tau)$, $i=2,3,4$.
The second equality in the third line of \eqn{theta fn} is known as the 
`Jacobi's triple product identity'.

The following identities are useful and repeatedly used in this paper:
\begin{eqnarray}
 && \sqrt{\frac{2\eta(\tau)}{\th_2(\tau)}} 
= \frac{\tTh{0}{1}(\tau)}{\eta(\tau)} 
\equiv \frac{1}{\eta(\tau)}\left(\Th{0}{4}(\tau)-\Th{4}{4}(\tau)\right),
\nn\\
&& \sqrt{\frac{\eta(\tau)}{\th_4(\tau)}} 
= \frac{\Th{1/2}{1}(\tau)}{\eta(\tau)} 
\equiv \frac{1}{\eta(\tau)}\left(\Th{1}{4}(\tau)+\Th{-3}{4}(\tau)\right),
\nn\\
&& \sqrt{\frac{\eta(\tau)}{\th_3(\tau)}} 
= \frac{\tTh{1/2}{1}(\tau)}{\eta(\tau)} 
\equiv
\frac{1}{\eta(\tau)}\left(\Th{1}{4}(\tau)-\Th{-3}{4}(\tau)\right).
\label{useful identity}
\end{eqnarray}
These are easily proved by using the Jacobi's triple product identity
as well as the Euler identity:
\begin{eqnarray}
 && 2\eta(\tau)^3 = \th_2(\tau)\th_3(\tau)\th_4(\tau) 
%\nn\\
\hspace{6mm}
 \left(\, 
\Longleftrightarrow~ \prod_{n=1}^{\infty}(1+q^{n})(1-q^{2n-1})=1 
\, \right).
\end{eqnarray}

%%%%%%%%%%%%%%%%%%%%%%%%%%%%%%%%%%%%%%%%%%%%%%%%%%%
%%%%%%%%%%%%%%%%%%%%%%%%%%%%%%%%%%%%%%%%%%%%%%%%%%%

%%%%% 
\section{Rational conformal models at $c=1$}
\label{sec:RCFT}
\setcounter{equation}{0}
\def\theequation{\ref{sec:RCFT}.\arabic{equation}}
%%%%%

Below we collect known facts about $c=1$ bosonic CFT which are instrumental in our T-fold analysis. 
When the compactification radius is $R=\sqrt{p/p'}$ ($p,p'$ are coprime positive integers) the 
bosonic system on $S^1$ or $S^1/{\B Z}_2$ exhibits an extended symmetry with respect to which the theory becomes rational. 
These symmetries are denoted ${\C A}_N$ (circle) or ${\C A}_N/{\B Z}_2$ (${\B Z}_2$-orbifold) in 
\cite{Dijkgraaf:1989hb}.
When $p=1$ or $p'=1$ the boundary states may be found by applying the Cardy's 
method\cite{Cardy:1989ir} as the rational CFT becomes diagonal.
 
\subsection{Rational Gaussian models}
The torus partition function of a boson $\varphi(z,\bar z)$ compactified on an $S^1$ at radius $R$ is
\bea
Z^{\msc{circ}}_R(\tau,\bar\tau)
&=&\frac{R}{\sqrt{{\rm Im}\tau}}\frac{1}{|\eta(\tau)|^2}
\sum_{m,w\in {\B Z}}\exp\left\{-\frac{\pi R^2|w\tau+m|^2}
{{\rm Im}\tau}\right\}\nn\\
&=&\frac{1}{|\eta(\tau)|^2}\sum_{k,\ell\in{\B Z}}
q^{\frac{1}{4}\left(\frac{k}{R}+R \ell\right)^2}
\bar q^{\frac{1}{4}(\frac{k}{R}-R\ell)^2}.
\eea
When the radius takes specific discrete values
\beq
R=\sqrt\frac{p}{p'}
\eeq
%(note: this is different from, but includes the cases of, rational multiples of the self-dual radius 
%$R=p\sqrt{\alpha'}/p$), 
there appears an extended algebra ${\C A}_N$ generated by operators of anomalous dimensions $h=1,N,N$,
\beq
j=  i\partial\varphi,\;\;\;
V^\pm=e^{\pm 2i \sqrt{N} \varphi},
\eeq
where
\beq
N=pp'
\eeq
(so $p\leftrightarrow p'$ gives the same chiral algebra, as it should).
At $N=1$ the ${\C A}_1$ is simply $SU(2)$ at level 1. 
There are $2N$ primary operators
\beq
\phi_k=e^{ik\varphi/{\sqrt{N}}}, \;\;\; k=0, 1,\cdots, 2N-1,
\eeq
whose conformal dimensions are
\beq
h_k=\min(\frac{k^2}{4N}, \frac{(2N-k)^2}{4N}).
\eeq
Corresponding character functions are
\beq
\chi_k(\tau)= \frac{\Th{k}{N}(\tau)}{\eta(\tau)}
\equiv \frac{1}{\eta(\tau)}\sum_{m\in{\B Z}}q^{(k+2mN)^2/4N}.
\label{eqn:circlechars}
\eeq
The partition function is written using the character functions as,
\beq
Z^{\msc{circ}}_R(\tau,\bar\tau)=\sum_{k=0}^{2N-1}\chi_k(\tau)\bar\chi_{\omega_0 k}(\bar\tau).
\label{eqn:circleZ}
\eeq
Here, $\omega_0$ is defined as 
\beq	
\omega_0=pr_0+p's_0\;\;\; ({\rm mod}\; 2N),
\eeq
using two integers $r_0$, $s_0$ satisfying $pr_0-p's_0=1\; ({\rm mod}\; 2N)$.
Such a pair (Bezout pair) $(r_0, s_0)$ is shown to be unique if restricted to region $1\leq r_0 \leq p'-1$, $1\leq s_0 \leq p-1$ and $p's_0<pr_0$.
The theory is diagonal when $p=1$ or $p'=1$.

The modular inversion of the ${\C A}_N$ characters is
\beq
\chi_k(-1/\tau)
=\sum_{\ell=0}^{2N-1}S_{k\ell}\chi_{\ell}(\tau)
=\frac{1}{\sqrt{2N}}\sum_{\ell=0}^{2N-1}e^{-i\pi k\ell/N}\chi_{\ell}(\tau),
\eeq
and the fusion rules are found by the Verlinde formula, 
\beq
\phi_i\times\phi_j=\sum_k N_{ij}^k\phi_k,\;\;\; N_{ij}^k=\delta_{i+j,k}.
\eeq
This simply reflects the conservation of the $U(1)$ charge.
The ${\C A}_N$ Ishibashi states $\vert\phi_\ell\rish$ are characterised by orthonormal overlaps 
\beq
\lish\phi_k\vert q^{\half(L_0+\overline{L}_0-\frac{1}{12})}\vert\phi_\ell\rish
=\delta_{k\ell}\chi_k(\tau).
\eeq
When $p=1$ or $p'=1$ there are $2N$ Cardy states 
that preserve the ${\C A}_N$ chiral symmetry, 
\beq
\vert\phi_k\rangle_C=\sum_{\ell=0}^{2N-1}\frac{S_{k\ell}}{\sqrt{S_{0\ell}}}\vert\phi_\ell\rish
=\frac{1}{\sqrt[4]{2N}}\sum_{\ell=0}^{2N-1}e^{-i\pi k\ell/N}\vert\phi_\ell\rish,
\label{eqn:AnCardy}
\eeq
where $S_{k\ell}$ is the modular inversion matrix.
In these cases the Cardy states are the Fourier transform of the Ishibashi states.
The inverse Fourier transformation is
\beq
\vert\phi_\ell\rish=(2N)^{-3/4}\sum_{k=0}^{2N-1}e^{i\pi k\ell/N}\vert\phi_k\rangle_C,
\eeq
where an obvious formula
$
\frac{1}{2N}\sum_{j=0}^{2N-1}e^{i\pi jk/N}=\delta_{k,0}^{(2N)}
$
has been used.
When $p'=1$ the $2N$ Cardy states may be identified with D-branes $\vert D(x_0)\rangle$ at $2N$ points on the circle, $x_0=0$, $\frac{\pi R}{N}$, $\frac{2\pi R}{N}$, $\cdots$, $\frac{(2N-1)\pi R}{N}$,
\beq
\vert\phi_k\rangle_C=\vert D(\frac{k\pi R}{N})\rangle
=\frac{1}{\sqrt[4]{2N^2}}\sum_{m\in{\B Z}}e^{-i\pi mk/N}\prod_{n=1}^\infty 
e^{\frac{a_{-n}\bar a_{-n}}{n}}\vert(m,0)\rangle,
\eeq
or Neumann states $\vert N(\tilde x_0)\rangle$ with $2N$ special values of the Wilson line, 
$\tilde x_0=0$, $\frac{\pi\alpha'}{RN}$, $\frac{2\pi\alpha'}{RN}$, $\cdots$, $\frac{(2N-1)\pi\alpha'}{RN}$ on the dual circle, 
\beq
\vert\phi_k\rangle_C=\vert N(\frac{k\pi\alpha'}{RN})\rangle
=\frac{1}{\sqrt[4]{2N^2}}\sum_{w\in{\B Z}}e^{-i\pi wk/N}\prod_{n=1}^\infty 
e^{-\frac{a_{-n}\bar a_{-n}}{n}}\vert(0,w)\rangle.
\eeq
%These are A-type boundary states of the $U(1)_N$ WZW theory \cite{Maldacena:2001ky}.
See \cite{Recknagel:1998ih,Gaberdiel:2001zq} for boundary deformation of these boundaries.

\subsection{Rational Gaussian orbifold models} 

The torus partition function of a boson compactified on an orbifold $S^1/{\B Z}_2$ at radius $R$ 
(i.e. on a line element of length $\pi R$) is
\beq
Z^{\msc{orb}}_R(\tau,\bar\tau)=\frac 12 
Z^{\msc{circ}}_R(\tau,\bar\tau)+\left|\frac{\eta(\tau)}{\theta_2(\tau)}\right|+\left|\frac{\eta(\tau)}{\theta_3(\tau)}\right|+\left|\frac{\eta(\tau)}{\theta_4(\tau)}\right|.
\eeq
The twisted part does not depend on the radius.
When $R^2=p/p'$ the CFT has an extended chiral symmetry ${\C A}_N/{\B Z}_2$, generated by 
($N=pp'$ as before)
\beq
T,\;\;\;
j_4=j^4-2j\partial^2 j+\frac 32 (\partial j)^2,\;\;\;
\cos(2 \sqrt{N} \varphi).
\eeq
Their conformal dimensions are $h=2,4,N$.
There are $N+7$ primary operators whose conformal dimensions are
\beq
\left.\begin{array}{ccccccccc}{\B I} & j & \phi^1_N & \phi^2_N & \phi_k & \sigma^1 & \sigma^2 & \tau^1 & \tau^2 \\ h=0 & 1 & N/4 & N/4 & k^2/4N & 1/16 & 1/16 & 9/16 & 9/16\end{array}\right.
\eeq
where $k=1,\cdots,N-1$.
Their character functions are
\bea
{\B I}:&&\chi_{\B I}(\tau)=\half\chi_0(\tau)+\frac{1}{2\eta(\tau)}\sum_{m\in{\B Z}}(-1)^mq^{n^2},\nn\\
j:&&\chi_{j}(\tau)=\half\chi_0(\tau)-\frac{1}{2\eta(\tau)}\sum_{m\in{\B Z}}(-1)^mq^{n^2},\nn\\
\phi^i_N:&&\chi^i_N(\tau)=\half\chi_N(\tau),\nn\\
\phi_k:&&\chi_k(\tau),\nn\\
\sigma^i:&&\chi_\sigma^i=\frac{1}{\eta(\tau)}\sum_{m\in{\B Z}}q^{(2m+\quarter)^2},\nn\\
\tau^i:&&\chi_\tau^i=\frac{1}{\eta(\tau)}\sum_{m\in{\B Z}}q^{(2m+\frac 54)^2},
\eea
where $i=1,2$ and $\chi_\ell(\tau)$ are the characters of the $S^1$ theory (\ref{eqn:circlechars}). 
The orbifold partition function at radius $R=\sqrt{p/p'}$ splits into 
the ${\C A}_N/{\B Z}_2$ characters,
\bea
Z^{\msc{orb}}_R(\tau,\bar\tau)
&=&|\chi_{\B I}(\tau)|^2+|\chi_j(\tau)|^2+|\chi^1_N(\tau)|^2+|\chi^2_N(\tau)|^2
+\sum_{k=1}^{N-1}\chi_k(\tau)\bar\chi_{\omega_0 k}(\bar\tau)\nn\\
&&+|\chi^1_\sigma(\tau)|^2+|\chi^2_\sigma(\tau)|^2+|\chi^1_\tau(\tau)|^2+|\chi^2_\tau(\tau)|^2.
\eea
Again the theory is not diagonal unless $p=1$ or $p'=1$.
The Cardy construction of boundary states in these diagonal cases is discussed for example in 
\cite{Oshikawa:1996dj,Hatzinikitas:2002ub}.
When $p=p'=1$ it turns out that the eight Dirichlet 
and Neumann states at the orbifold fixed points
\beq
D(0,\pm), \;\;\;
D(\pi R,\pm), \;\;\;
N(0,\pm), \;\;\;
N(\pi/R,\pm),
\eeq
may be identified %(in this order) 
with the ${\C A}_1/{\B Z}_2$ Cardy states
\beq
|{\B I}\rangle_C, \;\;\; |j\rangle_C, \;\;\;
|\phi^i_N\rangle_C, \;\;\;
|\sigma^i\rangle_C, \;\;\;
|\tau^i\rangle_C.
\eeq
%The remaining states $|\phi_k\rangle_C$ correspond to Dirichlet or
%Neumann states that are not on the fixed points.
Boundary deformation of the orbifold models is discussed e.g. in \cite{Recknagel:1998ih,Kawai:2006se}.

%%%%% 
\section{Twisted $SU(2)_k$ characters}
\label{sec:twistedSU(2)}
\setcounter{equation}{0}
\def\theequation{\ref{sec:twistedSU(2)}.\arabic{equation}}
%%%%%

In this Appendix 
we summarize formulae on the twisted characters of $SU(2)_k$.
We start by recalling  the 
%$\widehat{su(2)}_k$ 
$SU(2)_k$
character 
\beq
\chi^{(k)}_{\ell}(z |\tau)
\equiv \frac{\Theta_{\ell+1,k+2}(z|\tau)-\Theta_{-(\ell+1),k+2}(z|\tau)}
{i\theta_1(z|\tau)},
\label{eqn:SU(2)char}
\eeq
which is 
%(up to phase normalisation) 
a trace over the space of the spin 
$\ell/2$ module
%\footnote
%%
%{
%The normalization of $J^3$ here is $J^3(z)J^3(0) \sim \frac{k/2}{z^2}$.
%} 
%%
($0\leq \ell \leq k$),
\beq
\Tr_{{\cal H}^{(k)}_{\ell}} \left\lb 
q^{L_0-\frac{k}{8(k+2)}} e^{2\pi i z J^3_0}
\right\rb.
\eeq
Explicit forms of the $k=1$ and $k=2$ characters are
\begin{eqnarray}
 && \chi^{(1)}_{\ell}(z|\tau) =
  \frac{\Th{\ell}{1}(z|\tau)}{\eta(\tau)},~~ (\ell=0,1), 
\label{SU(2)_1 ch} \\
 && \chi^{(2)}_{0}(z|\tau) = \frac{1}{2} \left\lb 
\sqrt{\frac{\th_3(\tau)}{\eta(\tau)}}\frac{\th_3(z|\tau)}{\eta(\tau)}
+ \sqrt{\frac{\th_4(\tau)}{\eta(\tau)}}\frac{\th_4(z|\tau)}{\eta(\tau)}
\right\rb, \nn\\
&& \chi^{(2)}_{1}(z|\tau) = \sqrt{\frac{\th_2(\tau)}{2\eta(\tau)}} 
\frac{\th_2(z|\tau)}{\eta(\tau)}, \nn\\
&& \chi^{(2)}_{2}(z|\tau) = \frac{1}{2} \left\lb 
\sqrt{\frac{\th_3(\tau)}{\eta(\tau)}}\frac{\th_3(z|\tau)}{\eta(\tau)}
- \sqrt{\frac{\th_4(\tau)}{\eta(\tau)}}\frac{\th_4(z|\tau)}{\eta(\tau)}
\right\rb.
\label{SU(2)_2 ch}
\end{eqnarray}

%%%%%%%%%%%%%%%%%%%%%%%%%%%%%%%%%%%%%%%%%%%%%%%%%%%

We introduce the twisted characters 
by inserting operator $e^{2\pi i a J^3_0}$ along the spatial cycle and 
 $e^{2\pi i b J^3_0}$ along the temporal cycle of the world-sheet torus
($a,b\in {\B R}$).
Clearly the twist by the temporal insertion shifts the parameter $z$ by $b$. 
The twist in the spatial cycle may be taken into account by modular transformations.
With an appropriate choice of the phase normalisation 
the twisted characters are\footnote
{
There is phase ambiguity in defining the characters (see e.g. \cite{DiFrancesco:1997nk}) and the formula 
\eqn{eqn:twistedSU(2)char} 
is normalised so that they transform with the standard $SU(2)$ modular transformation laws.  
We normalise the twisted characters so that they behave in a modular covariant manner.
The choice is not unique; for instance the convention in \cite{Hellerman:2006tx} slightly differs from ours.
%and this is the reason why we can define another one $\chi^{(k)}_{\ell,[\alpha,\beta]}(\tau)$ 
% $(\alpha,\beta \in {\B Z}_2)$.
}
\beq
\widehat\chi^{(k)}_{\ell,(a,b)} %\left[\!\!\begin{array}{c} a\\ b\end{array}\!\!\right]
(z|\tau)
\equiv q^{\frac{k}{4}a^2} y^{\frac{k}{2}a}e^{2\pi i \frac{k}{4}ab}\,
\chi^{(k)}_{\ell} (z+a\tau+b|\tau).
\label{eqn:twistedSU(2)char}
\eeq
Their modular transformations are
\bea
&& \widehat\chi^{(k)}_{\ell,(a,b)}
%\left[\!\!\begin{array}{c} a\\ b\end{array}\!\!\right]
(z|\tau+1)
= e^{2\pi i(h_\ell-\frac{k}{8(k+2)})} 
\widehat\chi^{(k)}_{\ell,(a,b+a)}
%\left[\!\!\begin{array}{c} a\\ b+a\end{array}\!\!\right]
(z|\tau)
\label{eqn:T-twistedSU(2)char}, \\
&& \widehat\chi^{(k)}_{\ell,(a,b)}
%\left[\!\!\begin{array}{c} a\\ b\end{array}\!\!\right]
(\frac z\tau |\frac{-1}{\tau})
= e^{\frac{i\pi kz^2}{2\tau}} \,
\sum_{\ell'=0}^k S^{(k)}_{\ell,\ell'} \, 
\widehat\chi^{(k)}_{\ell',(b,-a)}
%\left[\!\!\begin{array}{c} b\\ -a\end{array}\!\!\right]
(z|\tau),
\label{eqn:S-twistedSU(2)char}
\eea
where 
$h_\ell=\frac{\ell(\ell+2)}{4(k+2)}$ is the conformal weights of the ground states and $
S^{(k)}_{\ell,\ell'}$ the modular $S$-matrix of $SU(2)_k$,
\begin{eqnarray}
 S^{(k)}_{\ell,\ell'} \equiv \sqrt{\frac{2}{k+2}} 
\sin \left(\pi\frac{(\ell+1)(\ell'+1)}{k+2}\right)~.
\end{eqnarray}

%%%%%%%%%%%%%%%%%%%%%%%%%%%%%%%%%%%%%%%%%%%%%%%%%%%

It is often  convenient to introduce the `${\B Z}_2$-twisted characters' 
$\chi^{(k)}_{\ell, [\alpha,\beta]}(\tau)$ 
whose boundary conditions are parameterized by 
${\B Z}_2$-valued indices $\alpha,\beta$. 
They are defined as 
\begin{eqnarray}
 && \chi^{(k)}_{\ell, [0,1]}(z|\tau) \equiv  
e^{\frac{i\pi}{2}\ell}\,
\widehat\chi^{(k)}_{\ell,(0,\half)}
%\left[\!\!\begin{array}{c} 0\\ \half\end{array}\!\!\right]
(z|\tau),\nn\\
 && \chi^{(k)}_{\ell, [1,0]}(\tau) \equiv 
 \widehat\chi^{(k)}_{\ell,(\half,0)}
%\left[\!\!\begin{array}{c} \half\\ 0\end{array}\!\!\right]
(z|\tau), \nn\\
 && \chi^{(k)}_{\ell, [1,1]}(z|\tau) \equiv 
 e^{-2\pi i \frac{k}{16}}e^{\frac{i\pi}{2}\ell}\,
\widehat\chi^{(k)}_{\ell,(\half,\half)}
%\left[\!\!\begin{array}{c} \half\\ \half\end{array}\!\!\right]
(z|\tau) 
%\nn\\
%&& 
\left(~ \equiv  
e^{2\pi i \frac{k}{16}}e^{-\frac{i\pi}{2}\ell}\,
\widehat\chi^{(k)}_{\ell,(\half,-\half)}
%\left[\!\!\begin{array}{c} \half\\ \half\end{array}\!\!\right]
(z|\tau)~ \right)
~.
\label{eqn:halftwistedSU(2)chars}
\end{eqnarray}
%%%%
%%%%
Their explicit forms using the theta functions are written as 
\begin{eqnarray}
\chi^{(k)}_{\ell,[0,1]}(z|\tau)& =& 
\frac{1}{\theta_2(z|\tau)} \left( 
\Th{-2(\ell+1)}{4(k+2)}(z/2|\tau) + (-1)^{\ell}\Th{2(\ell+1)}{4(k+2)}(z/2|\tau) 
\right.  \nn\\
&& ~~~ \left.
+ (-1)^k\Th{-2(\ell+1)+4(k+2)}{4(k+2)}(z/2|\tau) 
+ (-1)^{k+\ell}\Th{2(\ell+1)+4(k+2)}{4(k+2)}(z/2|\tau) 
\right) , \nn\\
%%%
%\left\{
%\begin{array}{ll}
%\dsp\frac{2}{\theta_2(z|\tau)} \left(
%\Theta_{2(\ell+1),4(k+2)}\left(\frac{z}{2}|\tau\right)
%+(-1)^k\Theta_{2(\ell+1)+4(k+2),4(k+2)}(\frac{z}{2}|
%\tau)
%\right)   &  ~~ (\ell~:~\mbox{even}) ,\\
%0 & ~~(\ell~:~\mbox{odd}).
%\end{array}
%\right.
%\nn\\
%%%
\chi^{(k)}_{\ell,[1,0]}(z|\tau)&=& 
\frac{1}{\theta_4(z|\tau)}\,
\left(\Theta_{-(\ell+1)+\frac{k+2}{2},k+2}(z|\tau)-
\Theta_{(\ell+1)+\frac{k+2}{2},k+2}(z|\tau)\right)  \nn\\
&\equiv & \frac{1}{\theta_4(z|\tau)}
\left( \Theta_{-2(\ell+1)+(k+2),4(k+2)}(z/2|\tau)
 - \Theta_{2(\ell+1)+(k+2),4(k+2)}(z/2|\tau) 
\right. \nn\\
 && \left. + \Theta_{-2(\ell+1)-3(k+2),4(k+2)}(z/2|\tau)
 - \Theta_{2(\ell+1)-3(k+2),4(k+2)}(z/2|\tau) \right) ,
\nn\\
%%%
\chi^{(k)}_{\ell,[1,1]}(z|\tau)&=& 
\frac{1}{\theta_3(z|\tau)}
\left( \Theta_{-2(\ell+1)+(k+2),4(k+2)}(z/2| \tau)
 +(-1)^{\ell} \Theta_{2(\ell+1)+(k+2),4(k+2)}(z/2|\tau) 
\right. \nn\\
 && \left. +(-1)^{k} \Theta_{-2(\ell+1)-3(k+2),4(k+2)}
(z/2|\tau)
 +(-1)^{k+\ell} \Theta_{2(\ell+1)-3(k+2),4(k+2)}
(z/2|\tau) \right). \nn \\
&&
\label{halftwistedSU(2)chars2-2}
\end{eqnarray}
Note that, when setting $z=0$, we have 
$\chi^{(k)}_{k-\ell,[1,0]}(0|\tau)=
\chi^{(k)}_{\ell,[1,0]}(0|\tau)$,  
$\chi^{(k)}_{k-\ell,[1,1]}(0|\tau)=
\chi^{(k)}_{\ell,[1,1]}(0|\tau)$,
and also 
$\chi^{(k)}_{\ell,[0,1]}(0|\tau)\equiv 0$ 
for an arbitrary odd $\ell$\footnote
 {
 These simple relations are broken when $z\neq 0$.
 }.

Taking level $k=1$ and setting $z=0$,
these characters reduce to the familiar conformal blocks of 
the twisted boson:
\begin{eqnarray}
 && \chi^{(1)}_{0,[0,1]}(0|\tau) 
= \frac{\widetilde\Theta_{0,1}(\tau)}{\eta(\tau)}
=\sqrt{\frac{2\eta(\tau)}{\theta_2(\tau)}}
~, ~~~
\chi^{(1)}_{1,[0,1]}(0|\tau)=0, \nn\\
&& \chi^{(1)}_{0,[1,0]}(0|\tau) =\chi^{(1)}_{1,[1,0]}(0|\tau) 
= \frac{\Theta_{1/2,1}(\tau)}{\eta(\tau)}
=\sqrt{\frac{\eta(\tau)}{\theta_4(\tau)}}, \nn\\
&& \chi^{(1)}_{0,[1,1]}(0|\tau) =\chi^{(1)}_{1,[1,1]}(0|\tau) 
= \frac{\widetilde\Theta_{1/2,1}(\tau)}{\eta(\tau)} 
=\sqrt{\frac{\eta(\tau)}{\theta_3(\tau)}}.
\label{twisted SU(2)_1 ch}
\end{eqnarray}
%%%%
Similarly, for $k=2$ we find the system of one twisted boson 
and one twisted fermion ($\th_i\equiv \th_i(0|\tau)$):
\begin{eqnarray}
&& \hspace{-1cm}
\chi^{(2)}_{0,[0,1]}(0|\tau) =
\frac{1}{2}\left(
\sqrt{\frac{\th_4}{\eta}}\frac{\th_3}{\eta}
+\sqrt{\frac{\th_3}{\eta}}\frac{\th_4}{\eta}
\right)=
\sqrt{\frac{2\eta}{\th_2}} \, 
\frac{1}{2}\left(\sqrt{\frac{\th_3}{\eta}}+
	    \sqrt{\frac{\th_4}{\eta}}\right), ~~~
\chi^{(2)}_{1,[0,1]}(0|\tau)=0, \nn\\
%%%
&& \hspace{-1cm}
\chi^{(2)}_{2,[0,1]}(0|\tau) = 
\frac{1}{2}\left(\sqrt{\frac{\th_4}{\eta}}\frac{\th_3}{\eta}-
\sqrt{\frac{\th_3}{\eta}}\frac{\th_4}{\eta}\right)=
\sqrt{\frac{2\eta}{\th_2}} \, 
\frac{1}{2}\left(\sqrt{\frac{\th_3}{\eta}}-
	    \sqrt{\frac{\th_4}{\eta}}\right), 
\nn\\
%%%
 &&
\hspace{-1cm}
 \chi^{(2)}_{0,[1,0]}(0|\tau) = \chi^{(2)}_{2,[1,0]}(0|\tau) =
\sqrt{\frac{\th_3}{\eta}}\frac{\th_2}{2\eta}=
\sqrt{\frac{\eta}{\th_4}}\sqrt{\frac{\th_2}{2\eta}}, ~~~
\chi^{(2)}_{1,[1,0]}(0|\tau) =
\sqrt{\frac{\th_2}{2\eta}}\frac{\th_3}{\eta}
= \sqrt{\frac{\eta}{\th_4}} \sqrt{\frac{\th_3}{\eta}},
\nn\\
%%%
&& \hspace{-1cm}
\chi^{(2)}_{0,[1,1]}(0|\tau)=\chi^{(2)}_{2,[1,1]}(0|\tau)=
\sqrt{\frac{\th_4}{\eta}}\frac{\th_2}{2\eta}
=\sqrt{\frac{\eta}{\th_3}}\sqrt{\frac{\th_2}{2\eta}}, ~~~
\chi^{(2)}_{1,[1,1]}(0|\tau)=
\sqrt{\frac{\th_2}{2\eta}}\frac{\th_4}{\eta}
= \sqrt{\frac{\eta}{\th_3}}\sqrt{\frac{\th_4}{\eta}}.
\label{twisted SU(2)_2 ch}
\end{eqnarray}

%%%%%%%%%%%%%%%%%%%%%%%%%%%%%%%%%%%%%%%%%%%%%%%%%%%%%%%%%%%%%%%%%%%%%%%%%

One can immediately see that 
the ground states 
of $\chi^{(k)}_{\ell,[0,1]}(z|\tau)$ are the usual spin 
$\ell/2$ integrable representation with conformal weights 
$h_{\ell}= \frac{\ell(\ell+2)}{4(k+2)}$.
On the other hand the ground states 
of $\chi^{(k)}_{\ell,[1,0]}(z|\tau)$ 
and $\chi^{(k)}_{\ell,[1,1]}(z|\tau)$ 
are the twisted sector vacuum whose conformal weight is
\begin{eqnarray}
h_{\ell}^{t} \equiv  
\frac{k-2+(k-2\ell)^2}{16(k+2)}+\frac{1}{16}
\equiv \frac{\ell(\ell+2)}{4(k+2)}-\frac{\ell}{4}
+ \frac{k}{16}
~.
\label{eqn:twistedh}
\end{eqnarray}

An important difference of the ${\B Z}_2$-twisted character
$\chi^{(k)}_{\ell, [0,1]}$ from $\hchi^{(k)}_{\ell,(0,1/2)}$ is that 
the insertion $e^{i\pi J^3_0}$ is now replaced with
$\widehat{\sigma} \equiv e^{i\pi \frac{\ell}{2}}e^{i\pi J^3_0}$. 
%(The operator `$e^{i\pi \frac{\ell}{2}}$' here just
%assigns the $\ell$-depending phase on the vacuum of spin
%$\ell/2$-representation, and is defined to be commuted with all the 
%$SU(2)$ currents.)
We note that $\widehat{\sigma}$ is involutive:
$\widehat{\sigma}^2= {\bf 1}$, whereas 
$e^{i\pi J^3_0}$ is not. 
The twisted characters of the other types  
$[\al,\beta]=[1,0],[1,1]$ are determined in a way consistent 
with the closedness of modular transformations. 
%($\Gamma(2)$-invariance) 
The modular transformations 
of $\chi^{(k)}_{\ell,[\al,\beta]}$
%of the ${\B Z}_2$ twisted characters 
are summarised as follows: 
\begin{eqnarray}
&&\hskip-20mm
\chi^{(k)}_{\ell,[0,1]}(z|\tau+1)= 
e^{2\pi i \left(h_\ell -\frac{k}{8(k+2)}\right)}\,
\chi^{(k)}_{\ell,[0,1]}(z|\tau)~, 
\hskip5mm
\chi^{(k)}_{\ell,[0,1]}\left(\frac{z}{\tau}|-\frac{1}{\tau}\right)
= e^{i\pi \frac{k}{2}\frac{z^2}{\tau}}\,
\sum_{\ell'=0}^k\, e^{\frac{i\pi}{2}\ell}
S_{\ell,\ell'}\,
\chi^{(k)}_{\ell',[1,0]}(z|\tau),  \nn\\
%%%
&&\hskip-20mm
\chi^{(k)}_{\ell,[1,0]}(z|\tau+1)=
e^{2\pi i\left(h^t_\ell-\frac{k}{8(k+2)}\right)}\, 
\chi^{(k)}_{\ell,[1,1]}(z|\tau)~, 
\hskip5mm
\chi^{(k)}_{\ell,[1,0]}\left(\frac{z}{\tau}|
-\frac{1}{\tau}\right)=
e^{i\pi \frac{k}{2}\frac{z^2}{\tau}}\,
\sum_{\ell'=0}^k\, S_{\ell,\ell'} e^{\frac{i\pi}{2}\ell'}\, 
\chi^{(k)}_{\ell'\,[0,1]}(z|\tau)~, \nn\\
%%%
&&\hskip-20mm 
\chi^{(k)}_{\ell,[1,1]}(z|\tau+1)= 
e^{2\pi i \left(h^t_\ell -\frac{k}{8(k+2)}\right)}\,
\chi^{(k)}_{\ell,[1,0]}(z|\tau)  ~, 
\hskip5mm
\chi^{(k)}_{\ell,[1,1]}\left(\frac{z}{\tau}|
-\frac{1}{\tau}\right)= e^{i\pi \frac{k}{2}\frac{z^2}{\tau}}\,
\sum_{\ell'=0}^k\, S_{\ell,\ell'}
e^{\frac{\pi i}{2}\left(\ell+\ell'-\frac{k}{2}\right)}\,
\chi^{(k)}_{\ell',[1,1]}(z|\tau) ~ . \nn\\
&&
\label{eqn:modulartwistedSU(2)-2} 
\end{eqnarray}
Note that $\widehat{\sigma}$ operates on the twisted Hilbert space ($\al=1$)
as
$\widehat{\sigma} \equiv e^{-i\frac{\pi}{4}k} e^{i\frac{\pi}{2}\ell} e^{i\pi J^3_0} $ 
that is again involutive\footnote
 {
This is easily checked using
$$
  \chi_{\ell,[\al,\beta]}(z+1|\tau) = e^{i\frac{\pi}{2}k \al}(-1)^{\ell} 
 \chi_{\ell,[\al,\beta]}(z|\tau).
$$
},
$\widehat{\sigma}^2= {\bf 1}$.
We may thus use the ${\B Z}_2$-twisted characters 
$\chi^{(k)}_{\ell,[\al,\beta]}$ as building blocks 
of the ${\B Z}_2$-orbifold of $SU(2)_k$.

Due to obvious global symmetry one may use 
$e^{i\pi J^1_0}$ or $e^{i\pi J^2_0}$ instead of $e^{i\pi J^3_0}$ above to define
the same twisted characters $\chi^{(k)}_{\ell,[\alpha,\beta]}(0|\tau)$.
%(with $e^{i\frac{\pi}{2}\ell}$ or $e^{-i\frac{\pi}{4}k}e^{i\frac{\pi}{2}\ell} $)
One can also use a more general rotated current zero mode 
$\rho e^{i\pi J^3_0} \rho^{-1}$, where $\rho$ is any automorphism of
$SU(2)$.
This is a consequence of the rotational invariance 
of the Hamiltonian and the property of trace. 
When the $U(1)$ dependence (the angle variable $z$) 
is turned on its zero-mode insertion 
must be rotated simultaneously, as $\rho e^{2\pi i z J^3_0} \rho^{-1}$.
We use these symmetries to compute various overlaps 
(see Appendix \ref{sec:MixedFormulae}).

%%%%%
\section{Formula for the Mixed Amplitudes}
\label{sec:MixedFormulae}
\setcounter{equation}{0}
\def\theequation{\ref{sec:MixedFormulae}.\arabic{equation}}
%%%%%

We derive in this Appendix the formula (\ref{eqn:formula1}) 
that was used in computing cylinder amplitudes of the fibre part.
Similar techniques were also utilized {e.g.} in \cite{Sugawara:2003xt}. 
%%%%%%%%%%%%%%%%%%%%%%%%%%%%%%%

We consider a boson compactified 
on a self-dual $S^1$ and let $\vert{N}\rangle$ be the 
Neumann boundary state,
\begin{eqnarray}
 \vert{N}\rangle = \frac{1}{2^{1/4}}\left(\vert{0}\rish+\vert{1}\rish\right)~.
\end{eqnarray}
Here $\vert{\ell}\rish$ are the $SU(2)_1$ Ishibashi states for the spin $\ell/2$ ($\ell = 0,1$) representations. 
These Ishibashi states are characterized by gluing conditions and overlaps,
\begin{eqnarray}
 && (J^a_n+\bar J^a_{-n})\vert{\ell}\rish=0~, ~~~
 \lish{\ell}\vert e^{-\pi s H^{c}} e^{2\pi i z J^3_0}\vert{\ell'}\rish = 
\delta_{\ell,\ell'}\frac{\Theta_{\ell,1}(z| is)}{\eta(is)} ~.
\label{Ishibashi ell}
\end{eqnarray}
It is then easy to find that  
\begin{eqnarray}
 \langle{N}\vert e^{-\pi s H^{c}} e^{2\pi i z J^3_0} \vert{N}\rangle
&=& \frac{1}{\sqrt{2}} \left(\frac{\Theta_{0,1}(z| is)}{\eta(is)}
+ \frac{\Theta_{1,1}(z| is)}{\eta(is)}\right) \nn \\
&=& \sum_{n\in {\B Z}} 
\frac{e^{-2\pi t \left(n+\frac{z}{2}\right)^2}}{\eta(it)} ~, ~~~
(t\equiv 1/s)~.
\label{eqn:formula0}
\end{eqnarray}

We wish to show that
\beq
\langle{N}\vert e^{-\pi s H^{(c)}} e^{2i\theta J^3_0} e^{2i\phi J^1_0}
\vert{N}\rangle = \frac{1}{\eta(it)} \sum_{n\in {\B Z}}
 e^{-2\pi t \left(n+\frac{\alpha(\theta,\phi)}{2\pi}\right)^2}~, 
\label{eqn:formula1}
\eeq
where
\beq
\alpha(\theta,\phi) \equiv \cos^{-1} 
\left(\cos\theta \cos \phi\right) ~.
\label{eqn:alphathetaphi}
\eeq
If this formula holds one may replace $J^1_0$ with $J^2_0$ because
$e^{-\frac{i\pi}{2}J^3_0} J^1_0 e^{\frac{i\pi}{2}J^3_0}= J^2_0$. 
%The formula \ref{formula} is derivable by the following 
%simple observation. 
One can show (\ref{eqn:formula1}) by going to the spin $\half$ basis of $SU(2)$ in which the 
current zero modes are represented by the Pauli matrices. 
Then one may write, 
\begin{eqnarray}
e^{2i\theta J^3_0} e^{2i\phi J^1_0}
= e^{i2\theta \frac{\sigma_3}{2}} e^{i2\phi \frac{\sigma_1}{2}}
= \left(
\begin{array}{cc}
 e^{i\theta}\cos \phi & ie^{i\theta}\sin \phi \\
 ie^{-i\theta} \sin \phi & e^{-i\theta}\cos \phi
\end{array}
\right)~.
\end{eqnarray}
This is diagonalised as
\begin{eqnarray}
 \left(
\begin{array}{cc}
 e^{i\alpha(\theta,\phi)}& 0 \\
 0 & e^{-i\alpha(\theta,\phi)}
\end{array}
\right) \equiv e^{2i \alpha(\theta,\phi) \frac{\sigma_3}{2}}~,
\end{eqnarray} 
with $\alpha(\theta,\phi)$ given by (\ref{eqn:alphathetaphi}).
We can then use an unitary operator $U$ to write
%\begin{eqnarray}
$ 
e^{2i\theta J^3_0} e^{2i\phi J^1_0} = U e^{2i\alpha(\theta,\phi)J^3_0}
U^{-1}, 
$
%\end{eqnarray}
where the explicit form of $U$ is
$U = e^{i\theta_1(J^{a_1}_0+\bar J^{a_1}_0)}
e^{i\theta_2(J^{a_2}_0+\bar J^{a_2}_0)} \cdots $. 
%Since we have $U^{-1}\vert{N}\rangle=\vert{N}\rangle$, we obtain the desired 
The Neumann state is invariant under the rotation by $U$
because of \eqn{Ishibashi ell}.
Therefore, using (\ref{eqn:formula0}), we obtain 
the desired formula (\ref{eqn:formula1}).
It is also easy to generalize the method described here to $SU(2)_k$
at arbitrary $k$.

\bigskip

\bigskip

%%%%%%%%%%%%%%%%%%%%%%%%%%%%%%%%%%%%%%%%%%%%%%%%%%%

%\bibliographystyle{JHEP}
%\bibliography{TfoldCFT}
%\input{TfoldCFT.bbl}
\providecommand{\href}[2]{#2}\begingroup\raggedright\endgroup

\end{document}